\documentclass{article}

\usepackage[margin=1.5in]{geometry}
\usepackage[utf8]{inputenc} 
\usepackage[T1]{fontenc}    
\usepackage{graphicx}       
\usepackage{subfig}         
\usepackage{tikz}           
\usepackage{multirow}
\usepackage{pgfplots}
\pgfplotsset{compat=1.17}

\graphicspath{ {./img/} }

\usetikzlibrary{shapes,decorations,arrows,calc,arrows.meta,fit,positioning,decorations.pathreplacing,decorations.pathmorphing}
\usepackage{apacite}
\usepackage{amsfonts}       
\usepackage{amsmath}        
\usepackage{mathtools}      
\usepackage{bm}             
\usepackage{microtype}      

\usepackage{calc}
\usepackage{float}
\usepackage{parskip}

\usepackage[ruled,vlined]{algorithm2e} 
\SetKw{Set}{set}
\SetKw{Ifthen}{If}
\SetKw{Cumsum}{cumulative sum}
\SetKw{Generate}{generate}
\SetKw{Sort}{sort}
\SetKw{Update}{update}
\SetKw{Save}{save}
\SetKw{myif}{if}
\SetKw{mythen}{then}
\SetKw{Return}{return}
\SetKwRepeat{myWhile}{while}{}

\tikzset{
    -Latex,auto,node distance =1 cm and 1 cm,semithick,
    state/.style ={ellipse, draw, minimum width = 0.7 cm},
    point/.style = {circle, draw, inner sep=0.04cm,fill,node contents={}},
    bidirected/.style={Latex-Latex,dashed},
    el/.style = {inner sep=2pt, align=left, sloped}
}

\newcommand\addvmargin[1]{
  \node[fit=(current bounding box),inner ysep=#1,inner xsep=0]{};
}

\title{A Bayesian semi-parametric approach for modeling memory decay in
dynamic social networks}

\author{Giuseppe Arena\thanks{Department of Methodology and Statistics, Tilburg School of Social and Behavioral Sciences,
Tilburg University, Warandelaan 2, 5037 AB Tilburg, The Netherlands}
 \and 
   Joris Mulder \footnotemark[1]\textsuperscript{\hspace{0.4em},}\thanks{Jheronimus Academy of Data Science, Sint Janssingel 92, 5211 DA ’s-Hertogenbosch, The Netherlands}
 \and
  Roger Th. A. J. Leenders \thanks{Department of Organization Studies, Tilburg School of Social and Behavioral Sciences,
  Tilburg University, Warandelaan 2, 5037AB Tilburg, The Netherlands}\textsuperscript{\hspace{0.4em},}\footnotemark[2]\\
}

\begin{document}
\maketitle
\begin{abstract}
In relational event networks, the tendency for actors to interact with each other depends greatly on the past interactions between the actors in a social network. Both the quantity of past interactions and the time that elapsed since the past interactions occurred affect the actors’ decision-making to interact with other actors in the network. Recently occurred events generally have a stronger influence on current interaction behavior than past events that occurred a long time ago--a phenomenon known as ``memory decay''. Previous studies either predefined a short-run and long-run memory or fixed a parametric exponential memory using a predefined half-life period. In real-life relational event networks however it is generally unknown how the memory of actors about the past events fades as time goes by. For this reason it is not recommendable to fix this in an ad hoc manner, but instead we should learn the shape of memory decay from the observed data. In this paper, a novel semi-parametric approach based on Bayesian Model Averaging is proposed for learning the shape of the memory decay without requiring any parametric assumptions. The method is applied to relational event history data among socio-political actors in India.
\end{abstract}


\section{Introduction}
\label{sec:introduction}
\par As a result of the growing automated collection of information,
fine-grained longitudinal network data are increasingly available in many
disciplines, such as sociology, psychology, and biology. These data have the potential to revolutionize our understanding about complex social network dynamics as we can learn how the past affects the future, how interaction behavior changes in continuous time, and how the actors' memory about past social interactions fades away as time progresses. This has stimulated social network
scientists to develop network models that suit the inherent dynamic nature of
these so-called \textit{relational event} data. A relational event is defined as an action initiated by a sender and targeted to one or more receivers at a specific point in time. The relational event modeling framework aims to model the \textit{event rate}, that is the speed at which relational events occur over a period of time between the actors in the model. The event rate can be expressed as a function of characteristics that quantify endogenous network patterns or exogenous information which determine how the network unfolds at some point in time \cite{Butts2008}.
In sociological and psychological research, the application of these relational event models aims to find behavioral patterns and to shed light on the emergence of a global
structure from network dynamics occurring at a local level \cite{Leenders2016}.
\par Of particular interest is to understand what triggers actors to interact with each other. Actors might decide which mutual recipient to target their actions to depending on various aspects such as homophily, norms of reciprocity, the volume of past social interactions, et cetera. Past relational events influence future events in different ways. First, this influence depends on qualitative aspects of the past events, such as whether the interaction was positive or a negative or who was the sender of the past event. For example, receiving a message from the company's president might have a greater effect than a message from another colleague. Events with a negative connotation have been argued to have a greater effect than events with a positive connotation. For instance a rebuke by the teacher may have a stronger effect than a praise by the teacher towards a child \cite{BrassLabianca}. Second, more recently past events generally have a greater influence on the present than events that occurred a long time ago \cite{Butts2008}. In a school setting a rebuke might have a longer lasting effect on future interactions than a praise and a rebuke coming from the teacher may have a longer-lasting effect than a rebuke coming from a classmate. This variability lays the foundation for the presence of an underlying memory process that shapes actors' decisions over time. The probability of remembering previously stored information has already been discussed in past studies where the memory retention concept was conceived as a stochastic psychological process. There, memory retention was studied by examining the dynamics of the hazard function corresponding to this stochastic process \cite{Chechile2003, Chechile2006,Chechile2009a}.
\par

In the relational event modeling literature, on the other hand, little attention has been
paid to the possibility that recently occurred events may have a larger impact on
the event rate than those events happened long ago. This is an important limitation as memory plays a crucial role in our understanding of social interaction between actors in a network. There are some noteworthy exceptions however. One approach has been to quantify a specific pattern of interactions according to two predefined different time interval definitions as in a 2-step approach: a \textit{short-term} expression (calculated by considering recently passed events) and a \textit{long-term} expression (considering only long passed events in the computation) \cite{Quintane2013}. The estimated effects of the two definitions describe how different the impact of the specific pattern is on the event rate according to different recency of events constituting the pattern itself. Another approach consists of estimating the model while using a moving time window of a specific memory length. The result is a discrete trend of the effects over the windows \cite{MulderLeenders2019}.
An alternative to time-intervals-based methods weighs events by means of an exponentially decreasing function with a given \textit{half-life} parameter that describes the elapsed time beyond which the weight of an event in the calculation of the statistic is halved \cite{Brandes2009}. A similar approach was suggested by \cite{Butts2008}. In all these cases we need to predefine the
\textit{memory lengths} both when defining short and long-term statistics and when choosing the memory length in the time window, or to a
particular \textit{steepness} of decay in the case of the half-life. The problem of the 2-step approach is that (i) it is unlikely that memory decays in a discrete step-wise manner, and (ii) the time points of the steps need to be chosen a priori (which is done in an ad hoc manner). The exponential decay on the other hand is modeled as a smooth function, which seems more natural. However the the half-life parameter which determines the steepness of the decay is arbitrarily chosen by the researcher a priori. The steepness of the decay however is generally unknown in practice.

\par The purpose of this work is to present a semi-parametric method for
learning the shape of memory decay in relational event models. The method is
semi-parametric in the sense that it does not make assumptions about a specific
functional form for memory decay. Indeed, parameters that potentially govern the memory process and, in turn, determine its shape over time are often unknown and our intent is to minimize the risk of mistakes that a parametric misspecification could induce. Our method
can be used for finding any functional form of memory decay which could be an
exponentially decreasing trend, a smoothed stepwise function, or other,
possibly more (or less) complex, functional trends.
In the case of event history data with
combinations of positive and negative events, the proposed method will not only
allow us to learn how much longer (or shorter) negative events are stored in the actor’s
memory than positive events but also whether the memory decay of these different
sentiments follow different functional shapes. Furthermore, this semi-parametric method combines the application of Bayesian inference in the context of a model selection problem (Bayesian Model Averaging) \cite{Volinsky1999} with the use of the relational event modeling framework as in \cite{Butts2008}.

\par The work is structured such that in Section 2 the relational modeling framework is introduced with a first insight over the concept of memory decay. In Section 3 a stepwise memory decay model is formulated. In Section 4 a continuous memory decay model is presented as well as the potential use of stepwise models in approximating the continuous shape of the decay. In Section 5 the semi-parametric method based on a Bayesian Model Averaging is presented along with two weighting systems used for generating random draws from the posterior memory decay. Finally, in Section 6 the method is applied to empirical data and findings are discussed, thus concluding with Section 7 where a few final considerations are taken around the methodology and its potential developments.

\section{Relational event models that capture memory decay}
\par In the relational event framework \cite{Butts2008}, a relational event $e_m$ is characterized by the 3-tuple $(s_{e_m},r_{e_m},t_{m})$, respectively sender, receiver and time of occurrence of the event. The joint probability of the realized ordered sequence of $M$ relational events,  $E_{t_M}=(e_1,\ldots,e_M)$, can be modeled as
\begin{equation}
    \begin{split}
    p(E_{t_M};\bm{\beta}) = 
                        \prod_{m=1}^{M}{\Bigg[
                        \lambda(s_{e_m},r_{e_m},X_{e_m},E_{t_{m-1}},\bm{\beta})
                        \prod_{e'\in\mathcal{R}}^{}{\exp{\left\lbrace - \lambda(s_{e'},r_{e'},X_{e'},E_{t_{m-1}},\bm{\beta})\left(t_m-t_{m-1}\right)\right\rbrace} }}
                        \Bigg]
    \end{split}
    \label{eq:joint_probability_rem}
\end{equation}
where $t_0$ (at $m=1$) is assumed equal to zero or to the starting time point of the case study,
$\lambda(s_{e_m},r_{e_m},X_{e_m},E_{t_{m-1}},\bm{\beta})$ is the rate of the event $e_m$ occurred at time $t_m$ and $\lambda(s_{e'},r_{e'},X_{e'},E_{t_{m-1}},\bm{\beta})$ represents the event rate of any event $e'$ that could have happened at time $t_m$ (including $e_m$). Indeed, $e'$ belongs to $\mathcal{R}$ 
which is the risk set consisting of all sender/receiver combinations, such as $\mathcal{S}\times R$: where $\mathcal{S}$ and $R$ are respectively sets of all possible senders and receivers for the entire event sequence. Where all actors can be senders as well as receivers in an interaction, then $\mathcal{S} \equiv R$ and the set of actors is simply indicated with $\mathcal{S}$.
Equation (\ref{eq:joint_probability_rem}) can be viewed as the well-known survival model with time-varying covariates, where hazard and survival components form the likelihood in the same way \cite{Lawless2003}.
\par The \textit{rate} of the specific dyadic event $e'\in\mathcal{R}$ at a generic time $t_m$ is modeled as a log-linear function of statistics and it can be written as follows
\begin{equation}
    \lambda(s_{e'},r_{e'},X_{e'},E_{t_{m-1}},\bm{\beta}) = \exp{\left\lbrace\sum_{p=1}^{P}{\beta_pu_p(s_{e'},r_{e'},X_{e'},E_{t_{m-1}})}\right\rbrace}
    \label{eq:event_rate_REM}
\end{equation}
where:
\begin{itemize}
    \item $\beta_p$ with $p=1,\ldots,P$, are parameters describing the effects of statistics on the logarithm of the event rate;
    \item $X_{e'}$ is the set of covariates (exogenous attributes, possibly time-varying) associated with event $e'$;
    \item $E_{t_{m-1}}$ refers to the collection of all of those events occurred before $t_m$;
    \item $u_p(s_{e'},r_{e'},X_{e'},E_{t_{m-1}})$ with $p=1,\ldots,P$, are the statistics of interest and each one can depend either on transpired events (endogenous statistics calculated at each time point, given $E_{t_{m-1}}$, and for either all dyads or all actors) or on exogenous attributes ($X_{e'}$).
\end{itemize}

\par In the standard specification of the model, endogenous statistics describe patterns of interactions occurring in the network which are quantified at each time point by considering the whole history of events happened from the initial state of the network (first observed relational event) until the time point before the current one (that is $t_{m-1}$ in (\ref{eq:event_rate_REM})). For instance, consider the standard formulation of the inertia statistic, which is a dyadic endogenous statistic that quantifies the volume of interactions of a specific dyad occurred until the current time point. Inertia quantifies the extent to which specific relational events keep repeating in the network over time. The corresponding formula at a generic time point $t_m$ with history $E_{t_{m-1}}$ will be the following,
\begin{equation}
    \text{inertia}(i,j,t_m)=\sum_{e\in E_{t_{m-1}}}^{}{\mathbb{I}_{e}(i,j)}
    \label{equation:inertia_standard_formula}
\end{equation}
\par where $\mathbb{I}_{e}(i,j)$ is the indicator variable that assumes value 1 if the event $e\in E_{t_{m-1}}$ has $s_{e} = i$ and $r_{e} = j$, 0 otherwise. The event rate for any possible event $e'\in\mathcal{R}$ at time $t_m$ with only the inertia in the linear predictor can be written as
\begin{equation}
    \lambda(s_{e'},r_{e'},E_{t_{m-1}},\bm{\beta}) = \exp{\left\lbrace \beta_{\text{inertia}} \text{inertia}(s_{e'},r_{e'},E_{t_{m-1}})\right\rbrace}
    \label{equation:event_rate_REM_with_inertia}
\end{equation}
A positive estimate for $\beta_{\text{inertia}}$ means that actors interact at higher rates with those actors who were often receivers of their past interactions. This is a sign of social routinization: what happened in the past is bound to be repeated over and over into the future,
For instance, consider Figure \ref{figure:inertia_example_with_intervals} where a sequence of events from $t_1$ to $t_{14}$ is represented on a time line. In order to calculate the inertia at time $t_{15}$ for the specific dyad $(i,j)$ we need to count the number of past events in the history $E_{t_{14}}$ where $i$ targeted an action to $j$, which is 6 in the example. Although this approach would give insights about how previous interactions between actors have influence on the event rate, we would be assuming long passed events (such as those that happened 14 and 11 events ago, over two hours ago) to be equally influential as recently passed ones (such as the ones that are only 1 or 4 events or 45 minutes or so old) in the computation of the statistics as well as on the event rate itself. This assumption may not be realistic for relational event data in practice as indicated earlier. Hence, the need of specifying a model that would be capable of accounting for this mutable effect by past events on dyadic event rate.

\section{A stepwise memory decay model}
\subsection{Stepwise decay for first order endogenous effects} 
\label{sec:A stepwise memory decay model:first order endogenous statistics}
\par 
We model the changing importance of past events as a function of the transpired time since the event was observed and do this through a stepwise memory decay model \cite{Perry2013}. First, the event history at each time point is divided into fixed intervals, then endogenous statistics are computed for each interval, and finally the corresponding endogenous effects are estimated. These effects quantify the relative importance of past events in predicting future events.
For instance, considering the event sequence in Figure \ref{figure:inertia_example_with_intervals}, we observe that at $t_{15}$ more than 2 hours transpired since the starting time point and we divide the history of events $E_{t_{14}}$ in three sub-histories according to a set $\bm{\gamma}$ of increasing time lengths, for example, $\bm{\gamma} = (0secs,30mins,2hrs,\infty)$
\begin{equation}
\begin{split}
    &E_{t_{14},1} = \left\lbrace e\in E_{t_{14}} : (t_{15}-t_e)\in(0secs,30mins] \right\rbrace\\
    &E_{t_{14},2} = \left\lbrace e\in E_{t_{14}} : (t_{15}-t_e)\in(30mins,2hrs] \right\rbrace\\
    &E_{t_{14},3} = \left\lbrace e\in E_{t_{14}} : (t_{15}-t_e)\in(2hrs,\infty) \right\rbrace\\
\end{split}
    \label{eq:sub-histories_definition}
\end{equation}
Where the first sub-history $E_{t_{14},1}$ contains all events transpired until 30 minutes before $t_{15}$; the second, $E_{t_{14},2}$, includes those events happened between 30 minutes and 2 hours before $t_{15}$; lastly, the third sub-history, $E_{t_{14},3}$, includes all events happened more than 2 hours before $t_{15}$ (the right bound is left undefined here).
In Figure \ref{figure:inertia_example_with_intervals}, the partition into sub-histories is shown by the upwards arrows corresponding to the time lengths $\bm{\gamma}$.
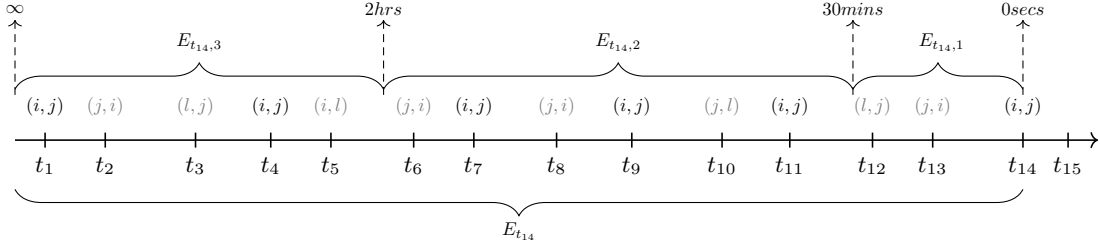
\begin{figure}[t!]
    \begin{center}
        \begin{tikzpicture}
            \draw[<-,densely dashed] (-7.4,1.6) -- (-7.4,0.55) node[below]{};
            \draw[<-,densely dashed] (-2.5,1.6) -- (-2.5,0.55) node[below]{};
            \draw[<-,densely dashed] (3.74,1.6) -- (3.74,0.55) node[below]{};
            \draw[<-,densely dashed] (6,1.6) -- (6,0.55) node[below]{};
            \draw[->,semithick] (-7.4,0) -- (7,0); 
            \draw [-,decorate,decoration={brace,amplitude=10pt},yshift=5pt] (-7.4,0.5) -- (-2.5,0.5) node [black,midway]{};
            \draw [-,decorate,decoration={brace,amplitude=10pt},yshift=5pt] (-2.5,0.5) -- (3.74,0.5) node [black,midway]{};
            \draw [-,decorate,decoration={brace,amplitude=10pt},yshift=5pt] (3.74,0.5) -- (6,0.5) node [black,midway]{};
            \draw [-,decorate,decoration={brace,amplitude=10pt},yshift=-30pt] (6,0.4) -- (-7.4,0.4) node [black,midway,yshift=-10pt,scale=0.70]{$E_{t_{14}}$};
            \foreach \x/\a in {-7/1,-6.2/2,-5/3,-4/4,-3.2/5,-2.1/6,-1.3/7,-0.2/8,0.8/9,2/10,2.9/11,4/12,4.8/13,6/14,6.6/15}
            \draw[-,semithick] (\x,0.1) -- (\x,-0.1) node[below]{\small$t_{\a}$};
            \node[scale=0.70] (x) at (-7,0.45) {$(i,j)$};
            \node[text=gray,scale=0.70] (x) at (-6.2,0.45) {$(j,i)$};
            \node[text=gray,scale=0.70] (x) at (-5,0.45) {$(l,j)$};
            \node[scale=0.70] (x) at (-4,0.45) {$(i,j)$};
            \node[text=gray,scale=0.70] (x) at (-3.2,0.45) {$(i,l)$};
            \node[text=gray,scale=0.70] (x) at (-2.1,0.45) {$(j,i)$};
            \node[scale=0.70] (x) at (-1.3,0.45) {$(i,j)$};
            \node[text=gray,scale=0.70] (x) at (-0.2,0.45) {$(j,i)$};
            \node[scale=0.70] (x) at (0.8,0.45) {$(i,j)$};
            \node[text=gray,scale=0.70] (x) at (2,0.45) {$(j,l)$};
            \node[scale=0.70] (x) at (2.9,0.45) {$(i,j)$};
            \node[text=gray,scale=0.70] (x) at (4,0.45) {$(l,j)$};
            \node[text=gray,scale=0.70] (x) at (4.8,0.45) {$(j,i)$};
            \node[scale=0.70] (x) at (6,0.45) {$(i,j)$};
            \node[scale=0.70] (x) at (-4.95,1.3) {$E_{t_{14},3}$};
            \node[scale=0.70] (x) at (0.62,1.3) {$E_{t_{14},2}$};
            \node[scale=0.70] (x) at (4.92,1.3) {$E_{t_{14},1}$};
            \node[scale=0.70] (x) at (6,1.75) {$0secs$};
            \node[scale=0.70] (x) at (3.74,1.75) {$30mins$};
            \node[scale=0.70] (x) at (-2.5,1.75) {$2hrs$};
            \node[scale=0.70] (x) at (-7.4,1.75) {$\infty$};
        \end{tikzpicture}
        \end{center}
        \caption{Example of inertia calculation relative to the dyadic event $(i,j)$ and given a sequence of events $t_1,\ldots,t_{14}$ with the corresponding history $E_{t_{14}}$.The event of interest in the calculation of the statistic is written in black, gray is otherwise. Without considering intervals, the value of inertia at time $t_{15}$ is 6. Whereas, when considering the interval definition (intervals delimited by upwards arrows) the value of inertia across the three intervals will become $\text{inertia}_1=1$, $\text{inertia}_2=3$ and $\text{inertia}_3=2$.}
        \label{figure:inertia_example_with_intervals}
\end{figure}
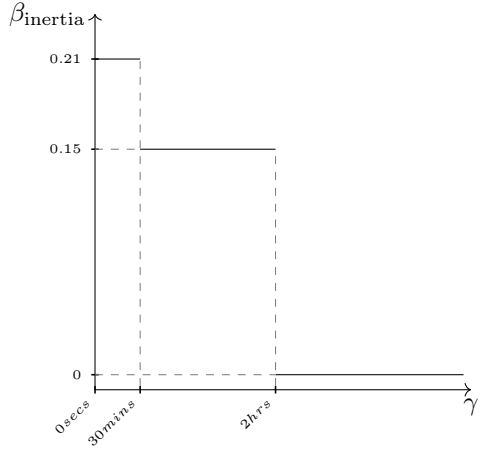
\begin{figure}[t!]
    \centering
    \begin{tikzpicture}
        \draw[->] (0,-0.2) -- (5,-0.2) node[below] {$\gamma$};
        \draw[->] (0,-0.2) -- (0,4.8) node[left] {$\beta_{\text{inertia}}$};
        \foreach \x/\t in {0/$0secs$,0.6/$30mins$,2.4/$2hrs$}
           \draw[-,semithick] (\x,-0.15) -- (\x,-0.25) node[left,rotate=45]{\tiny{\t}};
        \foreach \y/\b in {0/0,3/0.15,4.2/0.21}
           \draw[-,semithick] (0.05,\y) -- (-0.05,\y) node[left]{\tiny{\b}};
        \draw[-,domain=0:0.6,smooth,variable=\y,black]  plot ({\y},{4.2}); 
        \draw [-,dashed,gray] (0.6,-0.15) -- (0.6,4.2);
        \draw[-,domain=0.6:2.4,smooth,variable=\y,black]  plot ({\y},{3}); 
        \draw [-,dashed,gray] (0,3) -- (0.6,3);
        \draw [-,dashed,gray] (2.4,-0.15) -- (2.4,3);
        \draw[-,domain=2.4:4.9,smooth,variable=\y,black]  plot ({\y},{0}); 
        \draw [-,dashed,gray] (0,0) -- (2.4,0);
        \addvmargin{2mm}
    \end{tikzpicture}
    \caption{Stepwise effect of \textit{Inertia} on the event rate.}  
    \label{fig:dyadic_statistic_stepwise_effect_inertia}
\end{figure}
Therefore, three values of inertia can be calculated at any time point $t_m$ in the observed sequence by considering the three different partitions of the event history according to the increasing time lengths ($\bm{\gamma}$).
\begin{equation}
    \text{inertia}_k(i,j,t_{m}) = \sum_{e\in E_{t_{m-1},k}}^{}{\mathbb{I}_{e}(i,j)}\quad{}\text{with}\quad{}k=1,2,3
    \label{eq:inertia_in_intervals_example}
\end{equation}
Following the example, corresponding values of inertia according to intervals at time point $t_{15}$ are: $\text{inertia}_1=1$, $\text{inertia}_2=3$ and $\text{inertia}_3=2$.
We may expect that events occurred in $E_{t_{14},1}$ have a larger impact on the event rate than those occurred in $E_{t_{14},2}$ and $E_{t_{14},3}$. Although we do not make this assumption (as the goal is to learn from the data), the estimated effects relative to the three statistics will generally decrease in actual data, making the regression coefficient for inertia based on the most recent sub-history higher than that of inertia based on the most distant events, that is $\beta_{\text{inertia}_1}>\beta_{\text{inertia}_2}>\beta_{\text{inertia}_3}$.
 \par In a more general case where $K$ partitions of the current event history are defined according to increasing time lengths, such as
\begin{equation}
    \bm{\gamma} = (\gamma_0 , \gamma_1 , \ldots , \gamma_{K})\quad{\text{with}}
    \quad{}
    0=\gamma_0 < \gamma_1 < \ldots < \gamma_{K}=\infty
    \label{eq:widths_definition_general}
\end{equation}   
we can partition the event history $E_{t_{m-1}}$ at time $t_m$ into subsets as
\begin{equation}
    \begin{split}
    &E_{t_{m-1},1} = \left\lbrace e\in E_{t_{m-1}} : \gamma_{e}(t_m)\in(0,\gamma_1]
    \right\rbrace\\
    &E_{t_{m-1},2} = \left\lbrace e\in E_{t_{m-1}} : \gamma_{e}(t_m)\in(\gamma_1,\gamma_2]
    \right\rbrace\\
    &\vdots\\
    &E_{t_{m-1},K} = \left\lbrace e\in E_{t_{m-1}} : \gamma_{e}(t_m)\in(\gamma_{K-1},\infty) \right\rbrace\\
    \end{split}
    \label{eq:general_subhistory_definition}
\end{equation}
where $\gamma_{e}(t_m) = t_m - t_e$ represents the elapsed time at $t_m$ since the past event $e \in E_{t_{m-1}}$.
The general formula for inertia relative to the dyadic event $e$ with $(s_e=i,r_e=j)$ in the \textit{k}-th partition of the $E_{t_{m-1}}$ at time $t_m$ is
\begin{equation}
    \text{inertia}_k(i,j,t_m) = \sum_{e\in E_{t_{m-1},k}}^{}{\mathbb{I}_{e}(i,j)}\quad{}\text{with}\quad{}k=1,\ldots,K
    \label{eq:general_formula_inertia_in_intervals}
\end{equation}
The event rate for any possible event $e'\in\mathcal{R}$ at time $t_m$ where inertia is defined in $K$ partitions  will assume the form below,
\begin{equation}
    \lambda(s_{e'},r_{e'},E_{t_{m-1}},\bm{\beta}) = \exp{\left\lbrace \sum_{k=1}^{K}{\beta_{\text{inertia}_k}\text{inertia}_k(s_{e'},r_{e'},t_m)} \right\rbrace}
    \label{equation:event_rate_REM_with_interval_inertia}
\end{equation}
Once statistics are calculated across the $K$ partitions their corresponding parameters $\beta_{\text{inertia},k}$, with $k=1,\ldots,K$, can be estimated using the likelihood function in (\ref{eq:joint_probability_rem}). In the interval case for the inertia, parameters express how the propensity of actors in targeting their actions to the same past receivers changes as a function of the recency of past events themselves.
The use of interval statistics according to $K$ partitions of the event history, directly relates to the dynamic of the estimated effects and their evolution will follow a step function as in Figure \ref{fig:dyadic_statistic_stepwise_effect_inertia} with a mathematical function as in (\ref{equation:step_function_effects_elapsed_time}), that is based on the time lengths $\bm{\gamma}$ used to create the partitions.
\begin{equation}
    \beta_{\text{inertia}}(\gamma) = 
            \begin{cases}
                \beta_{\text{inertia}_1}& \text{if } \gamma\in(\gamma_0,\gamma_1]\\
                \vdots &\\
                \beta_{\text{inertia}_K}&\text{if } \gamma\in(\gamma_{K-1},\gamma_K]\\
                0              & \text{otherwise}
            \end{cases}
        \label{equation:step_function_effects_elapsed_time}
    \end{equation}
Stepwise memory effects can also be modeled for other first order endogenous statistics such as reciprocity, sender/receiver-in/out-degree whose formulas can be found in Appendix \ref{appendix:table_with_endogenous_statistics}.

\subsection{Stepwise decay for higher order endogenous effects} 
\label{subsec:stepwise_higher_order_endogenous_statistics}
Besides statistics that are based only on past interaction within a given dyad, the effects of higher order statistics involving more than two actors, can be used as well within this approach. Higher order endogenous statistics are characterized by more than one dyadic relational event in their formula. As such, the behavioral pattern of interest is more complex substantively as well as its computation. Indeed, in the case of triadic statistics, as with transitivity, the computation consists in the quantification of the number of times a dyad could potentially close a particular triangular structure if it occurred as next interaction after a specific sequence of past events.

\begin{figure}[t!]
    \centering
\subfloat[relational event $(i,l) \in E_{t_{m-1}}$ observed at time $t_m-\delta_1$.\label{fig:transitivity_stage_1}]{
    \begin{minipage}{0.30\textwidth}
    \centering
    \begin{tikzpicture}
        \node[state,scale=0.70] (x) at (0.1,1.9) {$i$};
        \node[state,scale=0.70] (y) at (-1.5,1) {$l$};
        \node[gray,state,scale=0.70] (z) at (0,0) {$j$};
        \node[above] (w) at (-0.75,2.2) {$t_m-\delta_1$};

        \path[scale=0.70] (x) edge[bend right,scale=0.70] (y);
        \addvmargin{2mm}
    \end{tikzpicture}
    \end{minipage}
    }
\hfill
\subfloat[relational event $(l,j) \in E_{t_{m-1}}$ observed at time $t_m-\delta_2$.\label{fig:transitivity_stage_2}]{%
  \centering
    \begin{minipage}{0.30\textwidth}
    \centering
    \begin{tikzpicture}
        \node[gray,state,scale=0.70] (x) at (0.1,1.9) {$i$};
        \node[state,scale=0.70] (y) at (-1.5,1) {$l$};
        \node[state,scale=0.70] (z) at (0,0) {$j$};
        \node[above] (w) at (-0.75,2.2) {$t_m-\delta_2$};

        \path[gray,scale=0.70,dashed] (x) edge[bend right,scale=0.70] (y);
        \path[scale=0.70] (y) edge[bend right,scale=0.70] (z);
        \addvmargin{2mm}
    \end{tikzpicture}
    \end{minipage}

  }
\hfill
\subfloat[relational event $(i,j)$ that can potentially happen at time $t_m$ closing then the triangular structure.\label{fig:transitivity_stage_3}]{%
  \begin{minipage}{0.30\textwidth}
  \centering
  \begin{tikzpicture}
      \node[state,scale=0.70] (x) at (0.1,1.9) {$i$};
      \node[gray,state,scale=0.70] (y) at (-1.5,1) {$l$};
      \node[state,scale=0.70] (z) at (0,0) {$j$};
      \node[above] (w) at (-0.75,2.2) {$t_m$};
  
      \path[gray,scale=0.70,dashed] (x) edge[bend right,scale=0.70] (y);
      \path[scale=0.70] (x) edge[bend left,scale=0.70] (z);
      \path[gray,scale=0.70,dashed] (y) edge[bend right,scale=0.70] (z);
      \addvmargin{2mm}
  \end{tikzpicture}
\end{minipage}
  }
  \caption{Figures from left to right describe the pattern of the transitivity closure in three steps. Given the event history $E_{t_{m-1}}$, the possible event $(i,j)$ occurring at $t_m$ (\ref{fig:transitivity_stage_3}) can close a triad already opened with a third actor ($l$ in the example) that acts as a broker in the process of information sharing/mediation. Events $(i,l)$ (\ref{fig:transitivity_stage_1}) and $(l,j)$ (\ref{fig:transitivity_stage_2}) occur by following the order in the example, where $\delta_1$ and $\delta_2$ at time $t_m$ are the transpired times since the two events $(i,l)$ and $(l,j)$, such that $0 \le\delta_2<\delta_1< t_m$. Therefore, the order of their occurrence is taken into account. Gray nodes and dashed gray arrows indicate respectively inactive actors and events already occurred, whereas active actors and the occurring dyadic event are in black.} 
  \label{fig:transitivity_graph_representation}   
\end{figure}
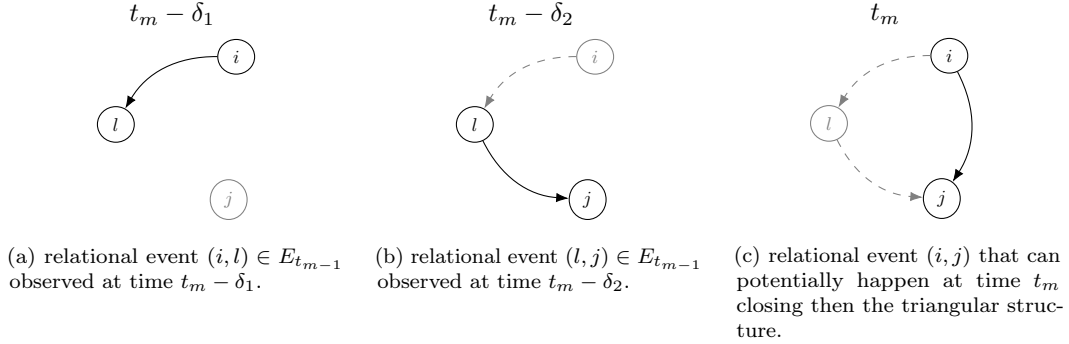

\par Figure \ref{fig:transitivity_graph_representation} describes the pattern of the transitivity closure in the context of relational event data where interactions are time-ordered. The search for specific behavioral patterns can be improved by introducing such time ordering in the calculation of the statistics. Specifically for the transitivity closure,  we designed the following formula that computes the statistic for the dyad $(i,j)$ at time $t_m$,
\begin{equation}
        \text{transitivity closure}(i,j,t_{m})=\sum_{l\in\mathcal{S}\setminus\left\lbrace i,j\right\rbrace}{
        \sum_{\substack{e\in E_{t_{m-1}}}}{\sum_{\substack{e^{*}\in E_{t_{m-1}}:\\ t_{e^{*}}\in[t_{e}-\gamma_{e}(t_m),t_{e})}}{\mathbb{I}_{e}(l,j)\mathbb{I}_{e^{*}}(i,l)}}}
    \label{eq:transitivity_new_formula}
\end{equation}
where:
\begin{itemize}
    \item $\mathbb{I}_{e}(l,j)$ is the indicator variable that assumes value 1 if the event $e\in E_{t_{m-1}}$ has $s_{e} = l$ and $r_{e} = j$, 0 otherwise (the same reasoning applies to the other indicator variables in (\ref{eq:transitivity_new_formula}));
    \item $e$ and $e^{*}$ are any pair of events belonging to the event history $E_{t_{m-1}}$ and such that $t_{e^{*}}<t_e$;
    \item $\gamma_{e}(t_m) = t_m-t_e$  is the time transpired at $t_m$ since the event $e\in E_{t_{m-1}}$.
\end{itemize}

\begin{figure}[t!]
    \begin{center}
        \begin{tikzpicture}
            \draw[->,semithick] (-7.8,0) -- (7,0); 
            \foreach \x/\a in {-6.2/2,-5/3,-4/4,-3.2/5,-2.1/6,-1.3/7,-0.2/8,0.8/9,2/10,3/11,3.8/12,4.4/13,5/14,6/15,6.6/16}
                \draw[-,semithick] (\x,0.1) -- (\x,-0.1) node[below]{};
            \node[text=gray,scale=0.70] (x) at (-6.2,0.3) {$(j,k)$};
            \node[text=red,scale=0.70] (x) at (-5,0.3) {$(i,l)$};
            \node[text=gray,scale=0.70] (x) at (-4,0.3) {$(k,l)$};
            \node[text=gray,scale=0.70] (x) at (-3.2,0.3) {$(j,k)$};
            \node[text=red,scale=0.70] (x) at (-2.1,0.3) {$(i,l)$};
            \node[text=gray,scale=0.70] (x) at (-1.3,0.3) {$(l,k)$};
            \node[scale=0.70] (x) at (-0.2,0.3) {$(l,j)$};
            \node[text=gray,scale=0.70] (x) at (0.8,0.3) {$(i,l)$};
            \node[text=gray,scale=0.70] (x) at (2,0.3) {$(j,k)$};
            \node[text=red,scale=0.70] (x) at (3,0.3) {$(i,l)$};
            \node[text=gray,scale=0.70] (x) at (3.8,0.3) {$(l,i)$};
            \node[scale=0.70] (x) at (4.4,0.3) {$(l,j)$};
            \node[text=gray,scale=0.70] (x) at (5,0.3) {$(l,k)$};
            \node[text=gray,scale=0.70] (x) at (6,0.3) {$(k,l)$};
            \node[scale=0.70] (x) at (-0.2,-0.3) {$t_{a}$};
            \node[scale=0.70] (x) at (4.4,-0.3) {$t_e$};
            \node[scale=0.70] (x) at (6,-0.3) {$t_{m-1}$};
            \node[scale=0.70] (x) at (6.6,-0.3) {$t_{m}$};
            \draw [-,decorate,decoration={brace,amplitude=10pt},yshift=-30pt] (6,-0.4) -- (-7.8,-0.4) node [black,midway,yshift=-10pt,scale=0.70]{$E_{t_{m-1}}$};
            \draw[-,yshift=15pt] (2.7,0)--++(90:0.2)--++(0:1.7)--++(-90:0.2);
            \draw[-,yshift=-15pt] (2.7,0)--++(-90:0.2)--++(0:1.7)--++(90:0.2); \node[scale=0.70] (x) at (3.55,-1.3) {$e^{*} : t_{e}-\gamma_{e}(t_m)\le t_{e^{*}}<t_{e}$};
            \draw[-] (2.7,-0.7) -- (2.7,0.7);
            \draw[-,yshift=15pt] (-7,0)--++(90:0.2)--++(0:6.785)--++(-90:0.2);
            \draw[-,yshift=-15pt] (-7,0)--++(-90:0.2)--++(0:6.785)--++(90:0.2);
            \node[scale=0.70] (x) at (-3.61,-1.3) {$a^{*}: t_{a}-\gamma_{a}(t_m)\le t_{a^{*}}<t_{a}$};
            \draw[-] (-7,-0.7) -- (-7,0.7);
            \draw[->] (-3.61,-0.76) -- (-3.61,-1.1) node[below]{};
            \draw[->] (3.55,-0.76) -- (3.55,-1.1) node[below]{};
            \node[state,scale=0.60] (x) at (-0.4,2.2) {$i$};
            \node[state,scale=0.60] (y) at (-0.9,1.4) {$l$};
            \node[state,scale=0.60] (z) at (0.1,1.4) {$j$};
            \path[scale=0.60] (x) edge[bend right,scale=0.60] (y);
            \path[scale=0.60] (y) edge[bend right,scale=0.60] (z);
        \end{tikzpicture}
        \end{center}
  \caption{Example of calculation of transitivity at $t_m$ for the dyad $(i,j)$ and information mediator $l$: the event history $E_{t_{m-1}}$ counts only two events $(l,j)$, at time $t_e$ and $t_a$. In order to quantify the contribute of $l$ to the $\text{transitivity}(i,j,t_m)$ we consider the intervals $[t_a-\gamma_{a}(t_m),t_a)$ and $[t_e-\gamma_{e}(t_m),t_e)]$ to seek for the first event in the pattern, that is $(i,l)$. The contribute of events $a$ and $e$ to the statistic are respectively 2 and 1. Then the value of transitivity for $(i,j)$ at $t_m$ with mediator $l$ is 3, meaning that if $(i,j)$ is the next event to occur it is going to close three potential triads where the information mediator was $l$.}
  \label{fig:transitivity_new_formula_tikz}   
\end{figure}
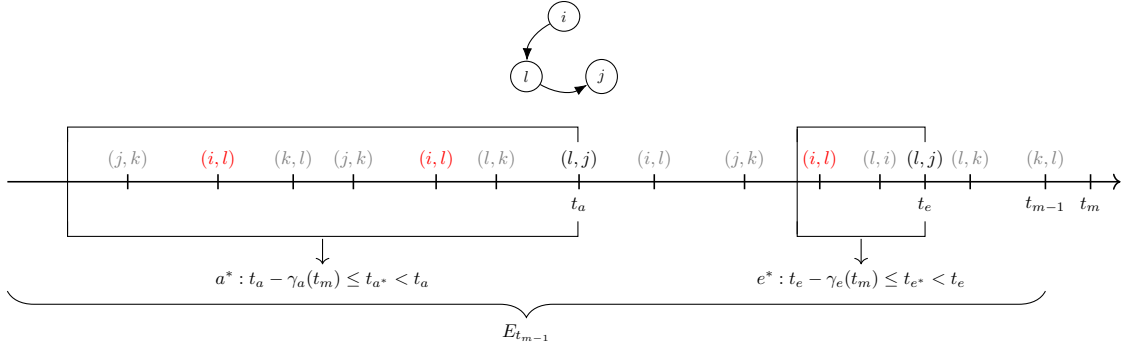

\par Figure \ref{fig:transitivity_new_formula_tikz} shows an example of the formula in (\ref{eq:transitivity_new_formula}) for just one $l \in \mathcal{S}\setminus\left\lbrace i,j\right\rbrace$ at time $t_m$, with a history of events $E_{t_{m-1}}$. In the example, two dyadic events $(l,j)$, noted as $e$ and $a$, occurred at $t_e$ and $t_a$ before $t_{m}$. For each of them we seek backward for those events $e^{*}$ and $a^{*}$ that occurred within intervals based on the transpired time of $e$ ($\gamma_{e}(t_m) = t_m-t_{e}$) and $a$ ($\gamma_{a}(t_m) = t_m-t_{a}$) which are respectively $[t_{e}-\gamma_{e}(t_m),t_{e})$ and $[t_{a}(t_m)-\gamma_{a},t_{a})$. Hence, if any event $e^{*}$ or $a^{*}$ in these intervals has sender $i$ and receiver $l$ then the product of the two indicator variables in (\ref{eq:transitivity_new_formula}) will be one and so will be contribute to the sum, 0 otherwise. 
In the specific example, as to event $a$ we observe two dyadic events $(i,l)$ happened in $[t_{a}-\gamma_{a}(t_m),t_{a})$, whereas for $e$ we find just one event $(i,l)$ occurred in $[t_{e}-\gamma_{e}(t_m),t_{e})$. Therefore, if the dyad $(i,j)$ is going to occur at $t_{m}$ it would close at least three potential triangular structures of the type in Figure \ref{fig:transitivity_graph_representation} where the actor $l$ is the information mediator.
The new formula for transitivity closure accounts for the time order of events in the triadic behavioral pattern and assumes that those events $(i,l)$ happened earlier than an event $(l,j)$ will count in the formula if and only if they transpired within the same time span of the specific $(l,j)$.
\par The event rate for any possible event $e'\in\mathcal{R}$ at time $t_m$ with only the transitivity in the linear predictor would then be written as
\begin{equation}
    \lambda(s_{e'},r_{e'},E_{t_{m-1}},\bm{\beta}) = \exp{\left\lbrace \beta_{\text{transitivity}} \text{transitivity}(s_{e'},r_{e'},E_{t_{m-1}})\right\rbrace}
    \label{eq:event_rate_REM_with_transitivity}
\end{equation}
A positive $\beta_{\text{transitivity}}$ means that the more partners $s_{e'}$ and $r_{e'}$ had in common in the past the more likely $s_{e'}$ will chose $r_{e'}$ as receiver of its coming interaction. Vice versa, when $\beta_{\text{transitivity}}<0$, the rate of the event $e'$ lowers, meaning that there is a tendency by actors in discouraging closure and thus, in fewer interactions with those actors they shared a partner with. However, the statistic in (\ref{eq:transitivity_new_formula}) refers to the event history $E_{t_{m-1}}$, that is the entire sequence of events since the onset until $t_{m-1}$ (including $e_{t_{m-1}}$).
Whereas the effect $\beta_{\text{transitivity}}$ may not be constant, depending on how recently the event $(l,j)$ occurred. Thus, transitivity can be re-defined in intervals in the same way as inertia in Section \ref{sec:A stepwise memory decay model:first order endogenous statistics}.
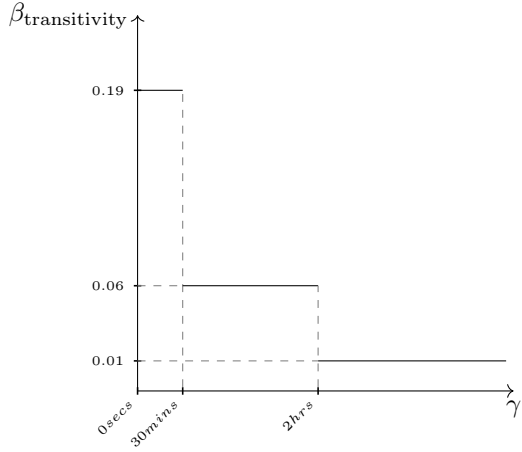
\begin{figure}[t!]
    \centering
    \begin{tikzpicture}
        \draw[->] (0,-0.2) -- (5,-0.2) node[below] {$\gamma$};
        \draw[->] (0,-0.2) -- (0,4.8) node[left] {$\beta_{\text{transitivity}}$};
        \foreach \x/\t in {0/$0secs$,0.6/$30mins$,2.4/$2hrs$}
           \draw[-,semithick] (\x,-0.15) -- (\x,-0.25) node[left,rotate=45]{\tiny{\t}};
        \foreach \y/\b in {0.2/0.01,1.2/0.06,3.8/0.19}
           \draw[-,semithick] (0.05,\y) -- (-0.05,\y) node[left]{\tiny{\b}};
        \draw[-,domain=0:0.6,smooth,variable=\y,black]  plot ({\y},{3.8}); 
        \draw [-,dashed,gray] (0.6,-0.15) -- (0.6,3.8);
        \draw[-,domain=0.6:2.4,smooth,variable=\y,black]  plot ({\y},{1.2}); 
        \draw [-,dashed,gray] (0,1.2) -- (0.6,1.2);
        \draw [-,dashed,gray] (2.4,-0.15) -- (2.4,1.2);
        \draw[-,domain=2.4:4.9,smooth,variable=\y,black]  plot ({\y},{0.2}); 
        \draw [-,dashed,gray] (0,0.2) -- (2.4,0.2);
        \addvmargin{2mm}
    \end{tikzpicture}
    \caption{Stepwise effect of \textit{Transitivity}  on the event rate.}
    \label{fig:triadic_statistic_transitivity_closure_in_intervals}  
\end{figure}

\par Consider the more general case of $K$ partitions of the current event history (as in (\ref{eq:general_subhistory_definition})) according to $K+1$ increasing time lengths $\bm{\gamma}$ (as in (\ref{eq:widths_definition_general})). The transitivity as regards the $k\text{-th}$ interval, for the dyad $(i,j)$ at time $t_m$ will be,
\begin{equation}
        \text{transitivity}_{k}(i,j,t_m)=\sum_{l\in\mathcal{S}\setminus\left\lbrace i,j\right\rbrace}{
        \sum_{\substack{e\in E_{t_{m-1},k}}}{\sum_{\substack{e^{*}\in E_{t_{m-1}}:\\ t_{e^{*}}\in[t_{e}-\gamma_{e}(t_m),t_{e})}}{\mathbb{I}_{e}(l,j)\mathbb{I}_{e^{*}}(i,l)}}}
    \label{eq:transitivity_interval_formula}
\end{equation}
where the quantification of potential triads is divided through the $K$ intervals of the history $E_{t_{m-1}} = \left\lbrace E_{t_{m-1},1},\ldots,E_{t_{m-1},K} \right\rbrace$ according to the time transpired at $t_m$ since the event $e$, that is $\gamma_{e}(t_m)$. However, the seeking of the event $e^{*}$ still considers the time interval as in (\ref{eq:transitivity_new_formula}). By using the interval formulation we are interested in understanding whether there exists an evolution of the transitivity effect on the event rate that depends on the recency of events constituting the triadic pattern. According to the stepwise formulation of the transitivity, we can rewrite the rate in (\ref{eq:event_rate_REM_with_transitivity})  as follows,
 \begin{equation}
    \lambda(s_{e'},r_{e'},E_{t_{m-1}},\bm{\beta}) = \exp{\left\lbrace\sum_{k=1}^{K}{\beta_{\text{transitivity}_k} \text{transitivity}_k(s_{e'},r_{e'},t_{m})}\right\rbrace}
    \label{equation:event_rate_REM_with_interval_transitivity}
\end{equation}
The effect of transitivity across intervals conveys more information than in the case without intervals. Although, the interpretation of positive and negative effects remains the same (i.e. positive effects still promote the closure of triads as well as negative effects keep discouraging it) the intensity of such behaviors that promote/discourage triadic closure can change over time and this one is the additional information we are after. For instance, if the effects from the first to the last interval are positive and decreasing, that is $\beta_{\text{transitivity}_1}>\ldots>\beta_{\text{transitivity}_K}$, this means that the estimated effects are encouraging the closure of triadic structures in such a way that the closer in time the events in the triad are to each other the faster the third event in the pattern is likely to happen.
\par The function in (\ref{equation:step_function_effects_elapsed_time}) can be written also in the case of triadic statistics and it will be,
\begin{equation}
    \beta_{\text{transitivity}}(\gamma) = 
            \begin{cases}
                \beta_{\text{transitivity}_1}& \text{if } \gamma\in(\gamma_0,\gamma_1]\\
                \vdots &\\
                \beta_{\text{transitivity}_K}&\text{if } \gamma\in(\gamma_{K-1},\gamma_K]\\
                0              & \text{otherwise}
            \end{cases}
    \label{equation:step_function_effects_elapsed_time_triadic_case}
\end{equation}
A simple example of stepwise effects as to the transitivity closure is shown in Figure \ref{fig:triadic_statistic_transitivity_closure_in_intervals}: if we only consider the transitivity closure in the model we can state that
the more the events characterizing the triad occurred in the recent times the more the third event in the triadic pattern is likely to happen. Formulas of further second order statistics can be found in Appendix \ref{appendix:table_with_endogenous_statistics}.

\subsection{Estimation of a relational event model with a stepwise memory decay}
The relational event model with stepwise memory decay of endogenous effects has the advantage that it can be easily estimated using existing software as \textit{relevent} \cite{Butts2008}, \textit{goldfish} \cite{Stadtfeld2020}, \textit{rem} \cite{Brandenberger2018}, \textit{remverse} \cite{Mulder2020}. This can be done as follows. First the transpired time needs to be divided in disjoint intervals with bounds $\gamma_0,\ldots,\gamma_K$. The bounds should be determined such that the stepwise function will be able to capture the expected memory. Thus for periods where a fast (slow) decay is expected narrow (wide) intervals should be chosen. Next, each endogenous statistic (e.g., inertia, transitivity) is split in $K$ separate statistics, which capture the volume of past interactions in the $K$ different intervals of the transpired time. The final set of relational event statistics can then be plugged into existing functions for fitting relational event models.

Despite the computational advantage, the stepwise memory decay in (\ref{equation:step_function_effects_elapsed_time}) and in (\ref{equation:step_function_effects_elapsed_time_triadic_case}) has two potential issues: a substantive issue is that it is unrealistic that memory decay occurs in a stepwise fashion in real life; a methodological issue is that it is unclear how many intervals ($K$) should be chosen and where the boundaries $\bm{\gamma} = (\gamma_0,\ldots,\gamma_K)$ should be placed.
Even though we could increase the number of intervals, we would still be constraining results to pre-specified boundaries and effects estimates would lose accuracy as this would greatly increase the number of free parameters in the model to be estimated and reduce the number of events per interval.

\section{The continuous nature of memory decay}
\label{sec:the_continuous_nature_of_memory_decay}
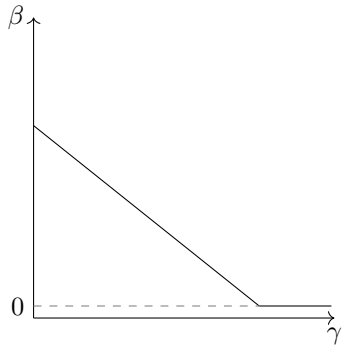
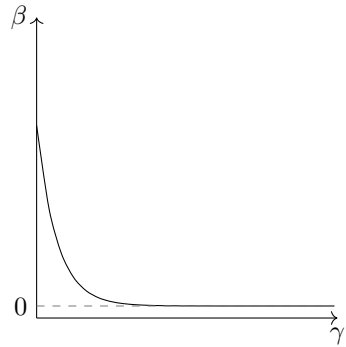
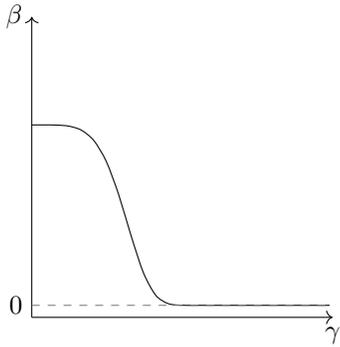
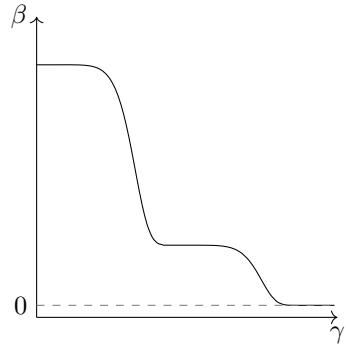
\begin{figure}[t!]
    \centering
\subfloat[linear decrease. \label{fig:effect_with_linear_decrease}]{
    \begin{minipage}{0.475\textwidth}
    \centering
    \begin{tikzpicture}[scale=0.8]
        \draw[->] (0,-0.2) -- (5,-0.2) node[below] {$\gamma$};
        \draw[->] (0,-0.2) -- (0,4.8) node[left] {$\beta$};
        \draw[-,domain=0:3.75,smooth,variable=\x,black] plot ({\x},{3-0.8*\x});
        \draw[-,domain=3.75:4.95,smooth,variable=\y,black]  plot ({\y},{0});
        \draw[-,domain=0:3.75,smooth,variable=\z,dashed,gray] node[left,black]{$0$}  plot ({\z},{0});
        \addvmargin{2mm}
    \end{tikzpicture}
    \end{minipage}
}
\hfill
\subfloat[exponential decrease.\label{fig:effect_with_exponential_decrease}]{%
  \centering
    \begin{minipage}{0.475\textwidth}
    \centering
    \begin{tikzpicture}[scale=0.8]
        \draw[->] (0,-0.2) -- (5,-0.2) node[below] {$\gamma$};
        \draw[->] (0,-0.2) -- (0,4.8) node[left] {$\beta$};
        \draw[-,domain=0:4.95,smooth,variable=\x,black] plot ({\x},{3*exp(-3*\x)});
        \draw[-,domain=0:4.95,smooth,variable=\z,dashed,gray] node[left,black]{$0$}  plot ({\z},{0});
        \addvmargin{2mm}
    \end{tikzpicture}
    \end{minipage}
}
\vskip\baselineskip
\subfloat[one-smooth-step decrease.\label{fig:effect_with_one_step_decrease}]{%
  \begin{minipage}{0.475\textwidth}
  \centering
  \begin{tikzpicture}[scale=0.8]
        \draw[->] (0,-0.2) -- (5,-0.2) node[below] {$\gamma$};
        \draw[->] (0,-0.2) -- (0,4.8) node[left] {$\beta$};
        \draw[-,domain=0:4.95,smooth,variable=\x,black] plot ({\x},{3*exp(-pow(0.6*\x,5))});
        \draw[-,domain=0:4.95,smooth,variable=\z,dashed,gray] node[left,black]{$0$}  plot ({\z},{0});
        \addvmargin{2mm}
  \end{tikzpicture}
\end{minipage}
}
\hfill
\subfloat[two-smooth-steps decrease.\label{fig:effect_with_two_steps_decrease}]{%
  \centering
    \begin{minipage}{0.475\textwidth}
    \centering
    \begin{tikzpicture}[scale=0.8]
        \draw[->] (0,-0.2) -- (5,-0.2) node[below] {$\gamma$};
        \draw[->] (0,-0.2) -- (0,4.8) node[left] {$\beta$};
        \draw[-,domain=0:2.1,smooth,variable=\x,black] plot ({\x},{1+3*exp(-pow(0.6*\x,8))});
        \draw[-,domain=2.1:4.95,smooth,variable=\x,black] plot ({\x},{exp(-pow(0.6*(\x-2.1),8))});
        \draw[-,domain=0:4.95,smooth,variable=\z,dashed,gray] node[left,black]{$0$}  plot ({\z},{0});
        \addvmargin{2mm}
    \end{tikzpicture}
    \end{minipage}
}
\caption{Possible evolution over $\gamma$ for the relative effect ($\beta$) of the past event constituting the endogenous statistic of interest. In these specific examples, trends decrease towards zero with different shapes depending on a set of parameters $\bm{\theta}$: (a) linear, (b) exponential, (c) an initial step with a smoothed decrease, (d) two smoothed and decreasing steps.} 
\label{fig:different_memory_evolutions}   
\end{figure}

\par 

Since memory decays in a smooth continuous trend by nature, we propose a more realistic form of the memory decay in (\ref{equation:step_function_effects_elapsed_time}) and (\ref{equation:step_function_effects_elapsed_time_triadic_case}) where, instead of constraining effects to be constant within intervals of $\gamma$, their change can be conceived as continuous over it and depend on a vector of parameters $\bm{\theta}$ which define the resulting shape of their decay. Therefore, the continuous effect for statistic $u$ can be written as
\begin{equation}
    \beta_{u}(\gamma,\bm{\theta}) 
        \label{eq:continuous_function_effects_elapsed_time}
    \end{equation}
where $\beta_u$ is now conceived as a continuous function on $\gamma$, describing the trend of the effect of $u$ such that $\beta_u : \mathcal{D}\rightarrow\mathbb{R}$ and $\mathcal{D} = \mathbb{R}^{+}\setminus\left\lbrace\gamma>\gamma_{K}\right\rbrace$, with $\gamma_K$ being a time length limit either due to the empirical data or simply justified by the researcher. The set of parameters $\bm{\theta} \in S(\bm{\theta})$ defines the shape of the decay, where $S(\bm{\theta})$ is their support.
We propose here several monotonous decreasing functions $\beta_u(\gamma,\bm{\theta})$ that might reflect the actual underlying memory decay. The continuous trends below assume effects to be positive and decreasing towards zero as the time transpired since the event increases.
\begin{itemize}
    \item linear decrease (Figure \ref{fig:effect_with_linear_decrease}):
        \begin{equation}
    \beta_u(\gamma,\theta_1,\theta_2) = \begin{cases}
                                \theta_2 - \frac{\theta_2}{\theta_1}\gamma & \text{for } \gamma<\theta_1\\
                                0 & \text{otherwise}
                                \end{cases}
        \label{eq:linear_decay_effect}
    \end{equation}
    where $\bm{\theta} = \left\lbrace\theta_1,\theta_2\right\rbrace$, $\theta_2 > 0$ is the maximum value assumed by the function and $-\frac{\theta_2}{\theta_1}$ (with $\theta_1 >0$) is the slope of the line which describes the steepness of the decrease; 
    \item exponential and one-smooth-step decrease (Figure \ref{fig:effect_with_exponential_decrease} and Figure \ref{fig:effect_with_one_step_decrease}):
    \begin{equation}
    \beta_u(\gamma,\theta_1,\theta_2,\theta_3) = \theta_3\exp{\left\lbrace-\left(\frac{\gamma}{\theta_1}\right)^{\theta_2}\right\rbrace}
    \label{eq:one_step_and_exponential_decay_effect}
    \end{equation}
    where the set of parameters  $\bm{\theta} = \left\lbrace\theta_1,\theta_2,\theta_3\right\rbrace$  consists of: $\theta_1 > 0$ and $\theta_3 > 0$ which are scale parameters ($\theta_3$ corresponds to the maximum value assumed by the function), $\theta_2 > 0$ is a shape parameter. The survival function of a Weibull distribution is a specific case of the function (\ref{eq:one_step_and_exponential_decay_effect}) where the maximum value is $\theta_3 = 1$. Moreover, where $\theta_2=1$, $\theta_3=\frac{1}{\theta_1}$, the (\ref{eq:one_step_and_exponential_decay_effect}) reduces to the exponential decreasing weight in \cite{Brandes2009} and the halflife parameter is then calculated as $T_{1/2}= \theta_{1}\log{2}$. The trend in most of the cases (except for the exponential one) starts evolving at an initial constant value (one-smooth-step trend) that is the maximum value $\theta_3$ and then decreases to zero as $\gamma$ increases;
    \item smoothed multiple steps (Figure \ref{fig:effect_with_two_steps_decrease}): it is a combination of two or more smoothed one-step trends.
\end{itemize}
The relative influence of past events on the dyadic event rate can follow other more complex shapes than those presented in Figure \ref{fig:different_memory_evolutions}.
As a result of this continuous definition of effects, inertia as well as other endogenous statistics are no longer computed as the accumulated number of past events but now consist of a sum of weights, where each weight changes according to the transpired time $\gamma$ of each event; this reflects the relative importance of past events updated at $t_{m}$. 
Therefore, the event rate in (\ref{equation:event_rate_REM_with_interval_inertia}) where only inertia effect is considered and inertia is divided in $K$ intervals becomes:
\begin{equation}
    \lambda(s_{e'},r_{e'},E_{t_{m-1}},\bm{\theta}) = \exp{\left\lbrace \sum_{e\in E_{t_{m-1}}}^{}{\mathbb{I}_{e}(i,j)\beta_{\text{inertia}}(\gamma_e(t_m),\bm{\theta})} \right\rbrace}
    \label{equation:2}
\end{equation}

\begin{figure}[htp]
    \centering
    \includegraphics[scale=0.5]{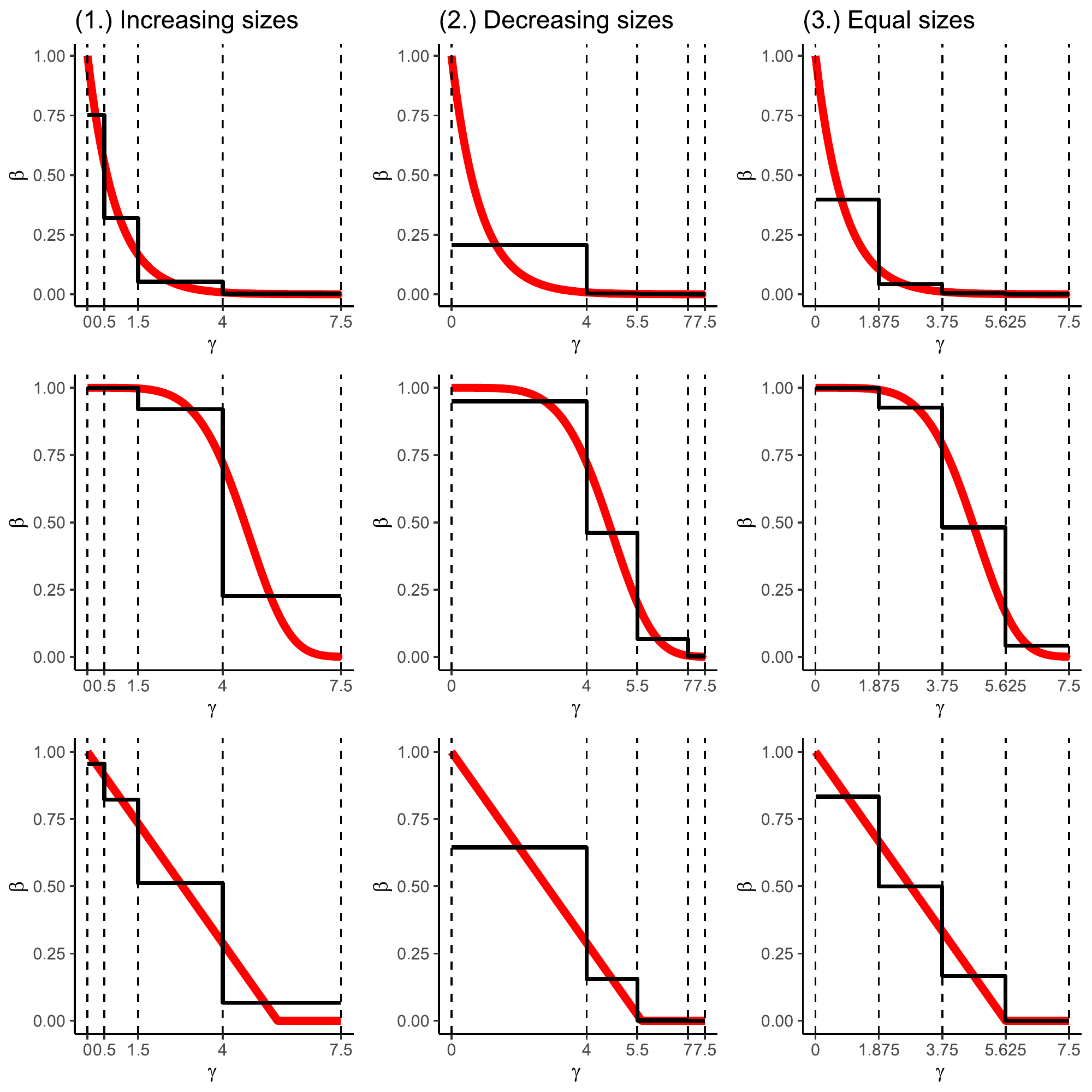}
  \caption{Examples of approximation of three different decays (red lines) by means of three types of stepwise functions (black lines). From the top to bottom: exponential decay, one-smooth-step decay and linear decay. From left to right: increasing sizes, decreasing sizes and equal sizes intervals. The maximum time width is $\gamma_{K}=7.5$.}
    \label{fig:example_method_widths}
\end{figure}
where $\beta(\gamma_{e}(t_m),\bm{\theta})$ is a continuous function that returns the relative effect as to the event $e$ contributing to the inertia statistics, $\gamma_{e}(t_m) = t_{m} - t_e$ is the time transpired at $t_{m}$ since $t_{e}$ and it increases over time, $\bm{\theta}$ is the set of parameters that describe the shape of the decay.
A formal mathematical procedure about moving from a stepwise effect function to a continuous effect function can be found in the Appendix \ref{appendix:from_stepwise_to_continuous_effects}.

\par However, the process of estimation of the set of parameters $\bm{\theta}$ governing the memory evolution results in a computationally complex maximization of the likelihood in (\ref{eq:joint_probability_rem}). The more realistic scenario that the influence of past events changes as a continuous function of their elapsed time since the current time, comes at the expense of constantly changing values of the network statistics, which increases the complexity of their estimation. By having stated and presented the continuous nature of the decay we also acknowledged the need of a less burdensome method to estimate parameters that govern memory. With such purpose, in the next subsection the use of a stepwise approach is revalued and presented in a Bayesian approach by which the posterior memory decay of effects is estimated.

\section{A semi-parametric approach to estimate a smooth memory decay}
    \label{sec:A semi-parametric approach to estimate memory evolution}
\par We introduce here a statistical approach based on the \textit{Bayesian Model Averaging} (BMA) \cite{Volinsky1999} which: (i) takes the computational advantage of the stepwise model introduced in Section (\ref{sec:A stepwise memory decay model:first order endogenous statistics}), (ii) avoids the issue of arbitrarily choosing intervals and (iii) results in a semi-continuous estimate of memory decay.

\subsection{Defining a bag of stepwise relational event models}
\label{subsec:defining a bag of stepwise relational event models}
The stepwise model we defined in the previous section pays off in the approximation of the memory decay, since the estimated effect $\beta_k$ represents the most likely relative effect that is assumed to be constant within the $k\text{-th}$ interval and given the sequence of $\bm{\gamma}$.
The level of approximation may depend both on the size and on the number of intervals. Therefore, a trade off between these two aspects can often result in a potentially satisfying approximation of the decay. If on one side the stepwise memory decay appears to be the least suitable model, on the other side we can take advantage of the approximation obtained by a set of different stepwise models and perform a Bayesian model averaging over the posterior trend of the effects. Such semi-parametric approach is introduced in the next section. At this stage, is necessary to describe a method for generating the sequences of time widths for each stepwise model that takes part in the final averaging.
The sequences of time widths may be generated according to three features reflecting three possible changes of the decay over time (a few examples are shown in Figure \ref{fig:example_method_widths}):
\begin{enumerate}
    \item[(i)] when memory change is likely to be stronger for the more recent events and to change less for events that already are in the farther past (where it is approximately constant) (e.g. exponential decay), then intervals with increasing sizes will better catch this behavior and their widths will follow the inequality:
    $\gamma_k-\gamma_{k-1} < \gamma_{k+1}-\gamma_{k}$ for $k=1,\ldots,K-1$. In other words, here memory is short such that events are "forgotten" fairly fast and the most recent events carry a much higher weight than somewhat less recent events, and fairly distant events have as little effect on the future as events from the far past.
    \item[(ii)] if the decay occurs in the long term memory (close to $\gamma_{K}$) whereas it is steady around a constant value during the more recent past (e.g. one-smoothed step decay), then intervals with decreasing sizes will be capable of catching this behavior and their widths will satisfy the inequality: $\gamma_k-\gamma_{k-1} > \gamma_{k+1}-\gamma_{k}$ for $k=1,\ldots,K-1$;
    This type of widths could be generated by simply inverting the increasing widths in (i). This represents the situation where the effect of events decays only slowly for a while, until they are far enough back in time, which is when they lose there effect fast (e.g., where events from the past week matter, but anything beyond that is quickly forgotten).
    \item[(iii)] if the decay is decreasing with a constant pace (e.g. linear decreasing function), intervals with the same size will better emulate this behavior.
\end{enumerate}
\par Increasing sizes intervals (i) are generated by means of an algorithm based on the Dirichlet distribution and its pseudocode can be found in Appendix \ref{appendix:intervals_generator}. Decreasing sizes intervals (ii) are generated by first drawing random intervals of the type (i) and then inverting the order of the widths.

\subsection{Bayesian model averaging for approximating smooth decay functions}
\par We  can take advantage of the information provided by a set of $Q$ stepwise models where each model $\mathcal{M}_{q}$, with $q=1,\ldots,Q$, will be based on predefined intervals of the elapsed time whose number and time lengths are different across models and are respectively noted as $K_q$ and $\bm{\gamma}_q = (\gamma_0,\gamma_1,\ldots,\gamma_{K_q})$ with $q=1,\ldots,Q$.
By means of BMA one can elicit a posterior estimate of a quantity of interest as well as its average posterior predictive distribution by finding the optimal linear combination of a set of models, and accounting, in turn, for their uncertainty. A crucial aspect in BMA is the use of model weights which quantify the relative importance of models according to their posterior probability.
The challenging aspect here is that the true model $\mathcal{M}^*$, which may have a smooth shape for memory decay, is not a stepwise model. In the literature this scenario is called an $\mathcal{M}\text{-open}$ problem \cite{Yao2018, Bernardo2000}. The idea however is that the true smooth model can be found by using a linear combination of stepwise models using appropriate weights $w_q$ using Bayesian model averaging, where $w_{q}$ represents the weight of the $q\text{-th}$ stepwise model.

\par Weighting systems for the posterior probabilities of stepwise models can be calculated using a measure that quantifies the goodness of fit (via the likelihood) or the goodness of prediction (via the prediction of future observations based on past observations). The former type comprises all of those measures that are functions of some Information Criterion, such as BIC (Bayesian Information Criterion). The latter refers to those measures which quantify predictive performance of the models, such as ELPD (Expected Log-pointwise Predictive Density). Therefore, in light of this distinction, we considered two weighting systems which can be employed in the estimation of the posterior trend: one that is the most common choice in BMA (and is based on the BIC) and one that is specifically recommended for$\mathcal{M}\text{-open}$ problems in other types of model selection problems (and is based on an approximation of the ELPD \cite{Watanabe2013,Vehtari2017,Yao2018}). Thereby we will explore which type of weighting system can best be used for this challenging model averaging problem.

\par In BMA, the posterior estimate of the parameter of interest $\psi$ can be calculated as the weighted mean of the posterior estimates provided by each model in the averaging,
\begin{equation}
p(\psi|E_{t_M}) = \sum_{q=1}^{Q}{p(\psi |\mathcal{M}_q,E_{t_M})w_{q}}
\label{eq:bma_general_formula_in_rem}
\end{equation}
\par
\begin{figure}[t!]
    \centering
\subfloat[posterior draws from the collection of (stepwise) models.\label{fig:step_1_2}]{

    \begin{minipage}{0.51\textwidth}
    \centering
    \includegraphics[scale=0.5]{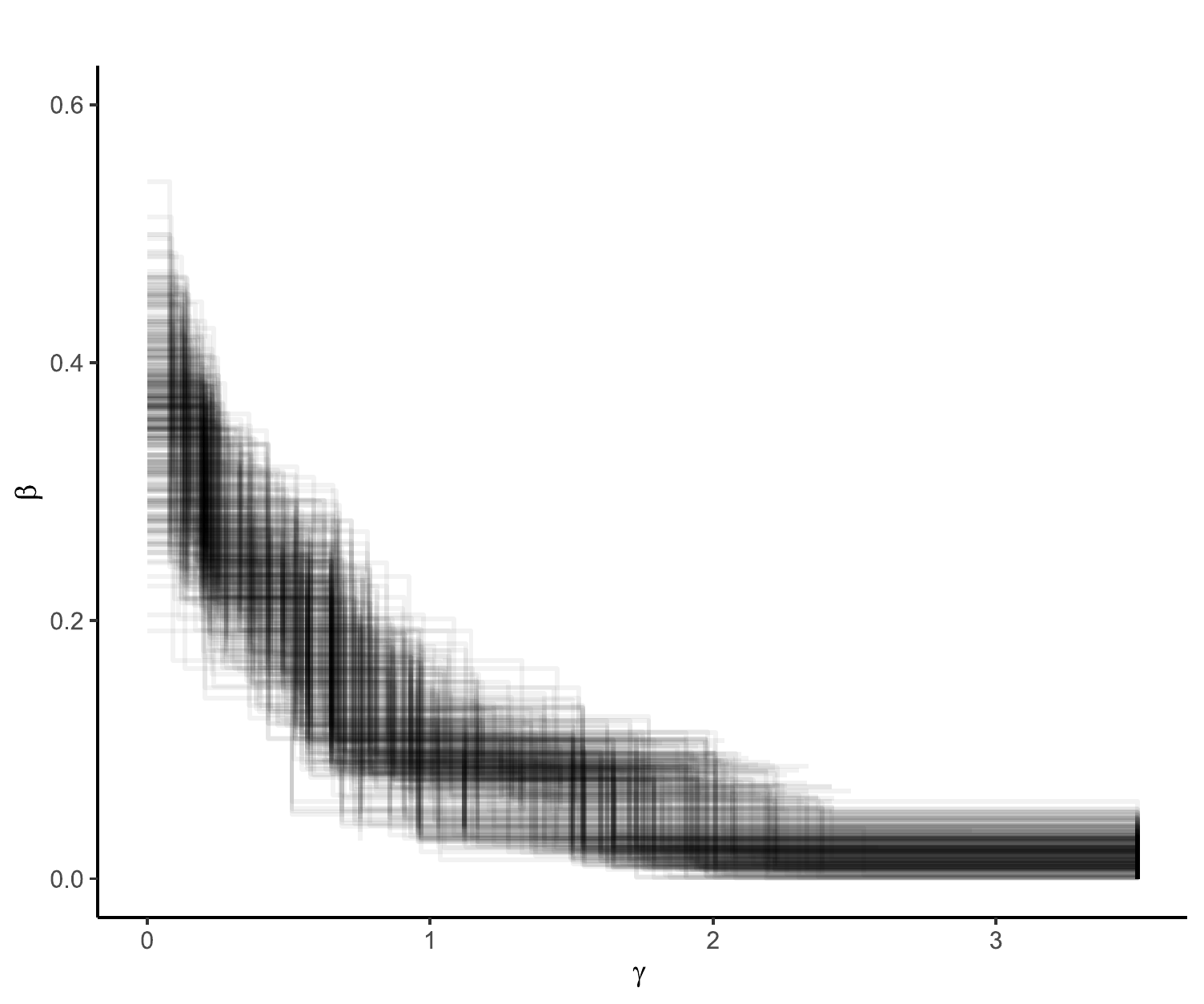}
    \end{minipage}
}
\subfloat[grid of $\gamma$'s.\label{fig:selecting_gammas}]{%
    \begin{minipage}{0.48\textwidth}
    \centering
    \includegraphics[scale=0.5]{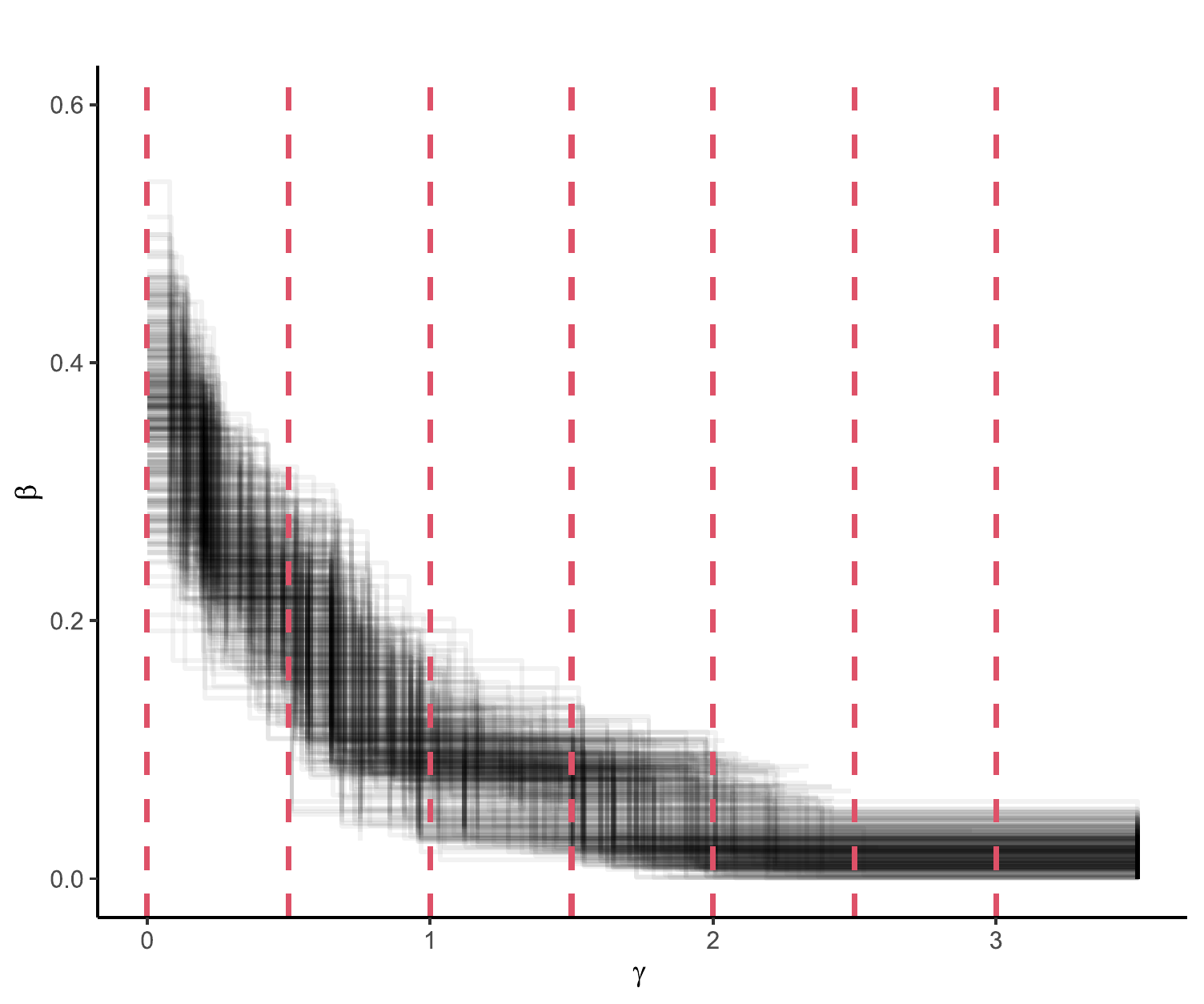}
    \end{minipage}
}
\vskip\baselineskip
\subfloat[posterior densities of $\beta$ at selected $\gamma$'s.\label{fig:getting_the_posterior_distribution_given_the_gamma}]{%
  \begin{minipage}{0.51\textwidth}
  \centering
     \includegraphics[scale=0.50]{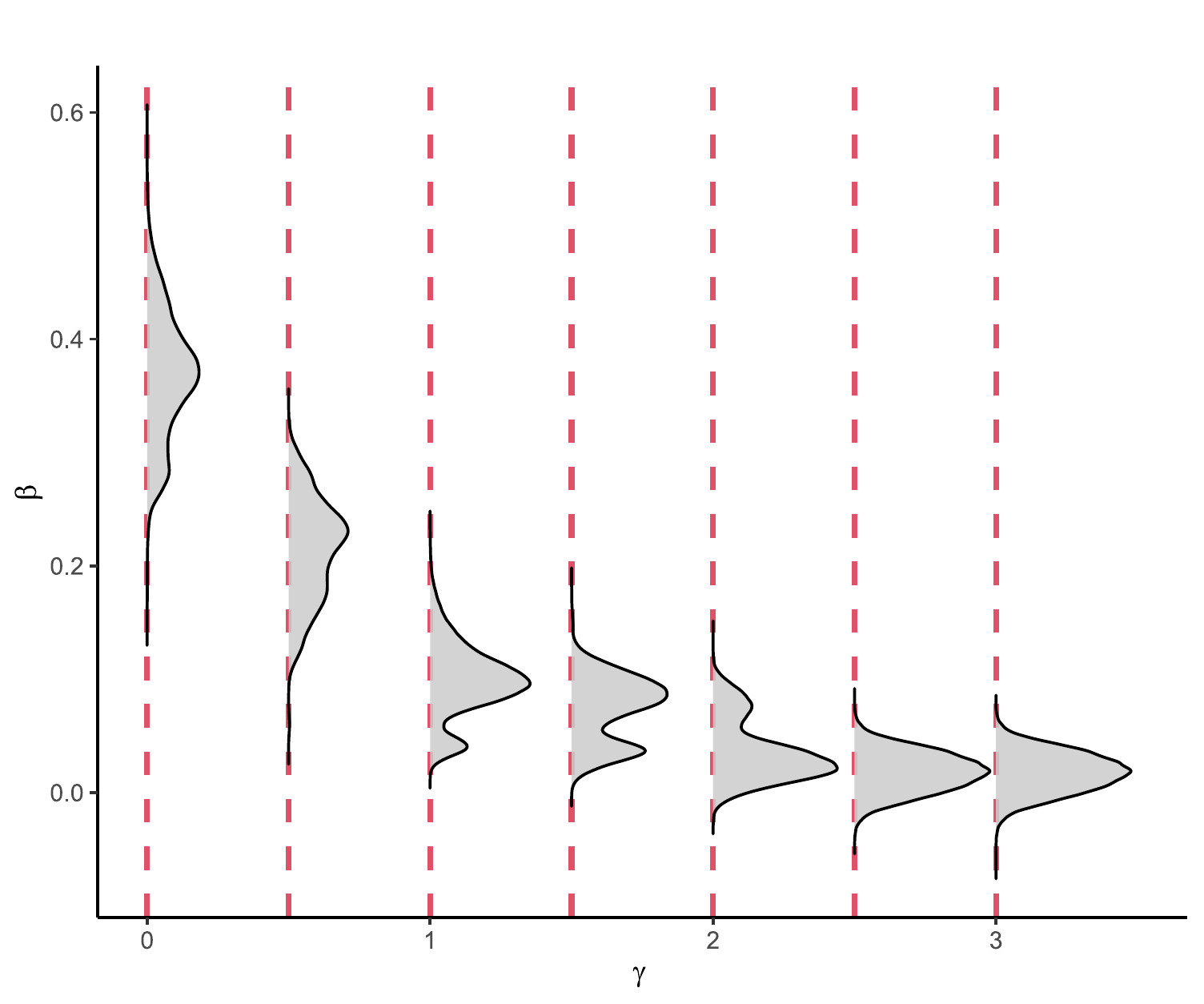}
\end{minipage}
}
\subfloat[resulting posterior trend of $\beta$ over a denser grid of $\gamma$'s.\label{fig:getting_the_posterior_distribution_given_more_gammas}]{%
  \begin{minipage}{0.48\textwidth}
  \centering
    \includegraphics[scale=0.5]{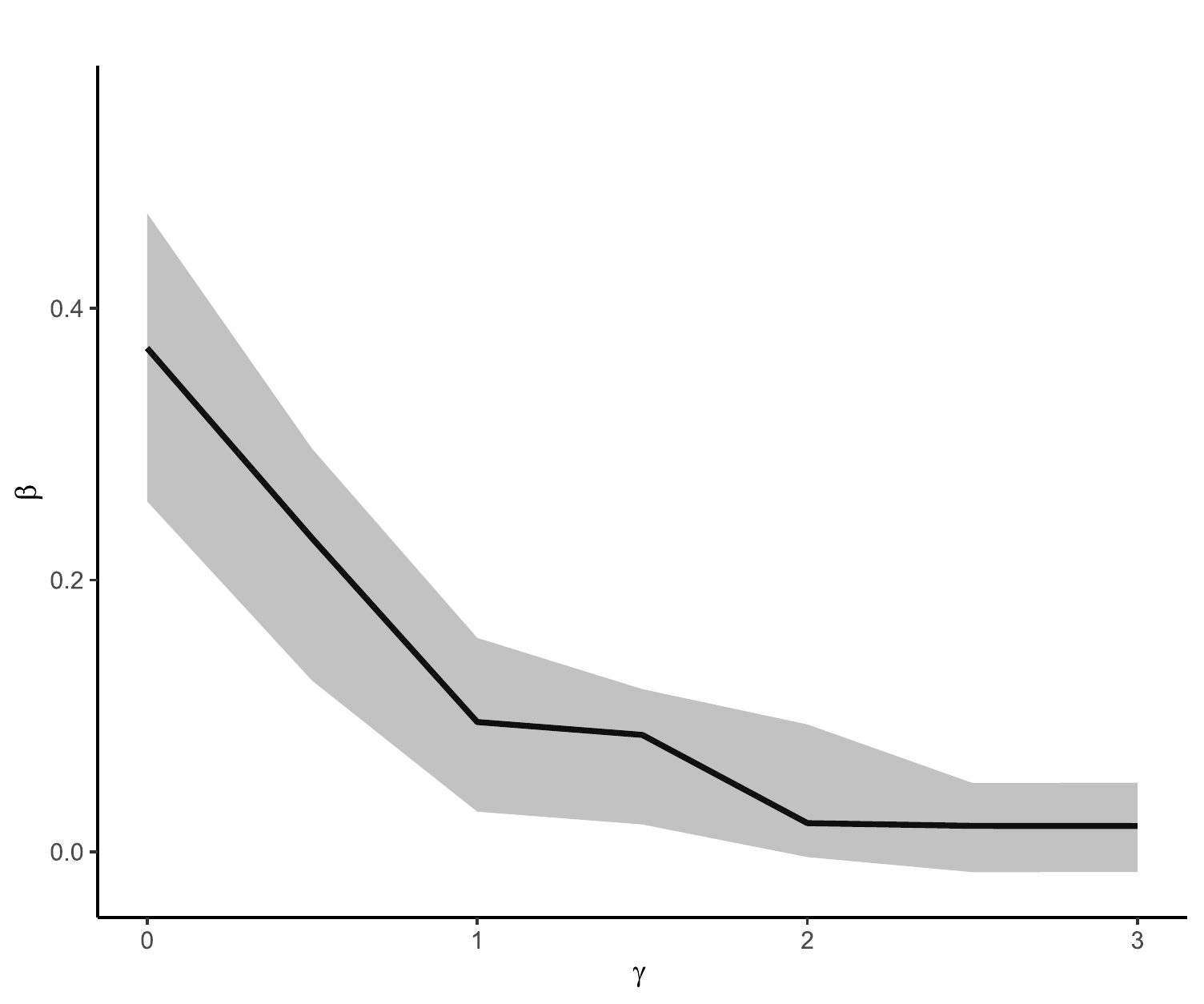}
\end{minipage}
}
\caption{(a) result from repeating step 1. and 2. in the approximation of the posterior distribution of $\beta$ over $\gamma$ by means of stepwise models;
(b) selecting a grid of $\gamma$'s (vertical dashed lines);
(c) posterior conditional densities at given $\gamma$'s;
(d) example of the posterior trend of $\beta$ over a denser grid of $\gamma$'s (the grey region represents the highest posterior density interval at 95\%).} 
\label{fig:posterior_distribution_of_effects_over_gamma}   
\end{figure}

 Considering (\ref{eq:bma_general_formula_in_rem}), it becomes easy to generate a posterior draw by first randomly selecting a model in the set and then generating a value from the posterior distribution of the selected model. In order to approximate the posterior distribution of $\bm{\beta}$ over $\gamma$ we first need to find the posterior distribution of $\bm{\beta}$ (without conditioning on any $\gamma$) and each draw is obtained with the three steps below:
\begin{enumerate}
    \item Draw a value from $q\sim\text{Multinomial}(\bm{w})$. This step consists in randomly choosing one of the $Q$ models according to the normalized vector of weights $\bm{w}=\left(w_{1},\ldots,w_{Q}\right)$ (weighting systems are discussed in \ref{subsec:BIC_weighting_system} and \ref{subsec:WAIC_weighting_system});
    \item Generate a vector of posterior effects from $\bm{\beta}|\mathcal{M}_q,E_{t_M}\sim MVN(\hat{\bm{\beta}}_{q},\hat{\bm{\Sigma}}_{q})$. The posterior distribution for the stepwise model $q$ (the model drawn at the first step) is approximated by a multivariate normal distribution with parameters given by maximum likelihood estimates of model $q$;
    \item Repeat steps 1 and 2 a sufficient number of times.
\end{enumerate}
After these three steps, the resulting posterior distribution of $\beta$ over $\gamma$ looks like the one in Figure \ref{fig:step_1_2}. Afterwards, the posterior decay of the effect over $\gamma$ is estimated as follows: (i) a (dense) grid is defined with evenly spaced $\gamma\in[0,\gamma_{max}]$, where $\gamma_{max}$ is usually based on the data (Figure \ref{fig:selecting_gammas}); (ii) for each $\gamma$ the corresponding interval effect in each posterior draw is selected (as it is shown by stepwise functions in (\ref{equation:step_function_effects_elapsed_time}) and (\ref{equation:step_function_effects_elapsed_time_triadic_case})), therefore this selection will result in a posterior density at a given $\gamma$ (Figure \ref{fig:getting_the_posterior_distribution_given_the_gamma}); (iii) the posterior mode of these densities is calculated at each $\gamma$, resulting in a semi-continuous effect decay (Figure \ref{fig:getting_the_posterior_distribution_given_more_gammas}).
As a consequence of this, the posterior estimate of those statistics that are not defined in intervals (e.g, baseline effect) is simply obtained with the draws generated after the three initial steps.

\subsection{Weighting systems for Bayesian model averaging}

\subsubsection{BIC weights}
\label{subsec:BIC_weighting_system}
When performing BMA, each model in the set $\mathcal{M}$ is considered as a generative model and it is weighted according to its posterior probability.
\begin{equation}
    p(\mathcal{M}_q|E_{t_M})=\frac{p(E_{t_M}|\mathcal{M}_q)p(\mathcal{M}_q)}
    {\sum_{q=1}^{Q}{p(E_{t_M}|\mathcal{M}_q)p(\mathcal{M}_q)}}
    \label{eq:posterior_probability_bma}
\end{equation}
Weights, in turn, depend on the marginal likelihood under each model, that is 
\begin{equation}
    p(E_{t_M}|\mathcal{M}_q)=\int_{}^{}{\cdots\int_{}^ {}{p(E_{t_M}|\bm{\beta}_{q},\mathcal{M}_q)p(\bm{\beta}_{q}|\mathcal{M}_q)d\bm{\beta}_{q}}}
    \label{eq:marginal_likelihood_bma_formula}
\end{equation} 
The marginal likelihood $p(E_{t_M}|\mathcal{M}_q)$ is sensitive to the prior assumption $p(\bm{\beta}_{q} | \mathcal{M}_q)$ of each model where, for instance, parameters $\bm{\beta}_{q}$ can be assumed mutually independent.

\par The marginal likelihoods of models can be approximated by means of a function of the Bayesian Information Criterion (BIC). The formula to calculate the unnormalized weight of a model in the set is,
    \begin{equation}
        p(\mathcal{M}_q|E_{t_M})\propto p(E_{t_M}|\mathcal{M}_q)p(\mathcal{M}_q)\approx \exp{\left\lbrace-BIC_{q}/2\right\rbrace}p(\mathcal{M}_q)
        \label{eq:bic_formula_result}
    \end{equation} 
where the prior distribution for each model is assumed to be non-informative $p(\mathcal{M}_q)=1/Q$, with $q=1,\ldots,Q$, because time lengths and number of intervals characterizing each stepwise model were randomly generated. Moreover, the marginal likelihood $p(E_{t_M}|\mathcal{M}_q)$ is approximated (with error $\mathcal{O}(1)$) by a function of the BIC of the $q\text{-th}$ model \cite{Konishi2008}.
    \par Thus, the normalized BIC weight for the $q\text{-th}$ model will be,
    \begin{equation}
        w^{\text{BIC}}_q = \frac{\exp{\left\lbrace-BIC_{q}/2\right\rbrace}}{\sum_{r=1}^{Q}{\exp{\left\lbrace-BIC_{r}/2\right\rbrace}}}
        \label{eq:bic_normalized_weight}
    \end{equation} 
Resulting weights are based on a fitting measure and the model that best fits the data will have the highest influence in estimating posterior distribution of interest, resulting in a poor approximation of the posterior trend $\bm{\beta}(\gamma)$. Furthermore, in the $\mathcal{M}\text{-open}$ case the use of posterior model probabilities is no longer suitable since the true model is out of the set of models.

\subsubsection{WAIC weights}
\label{subsec:WAIC_weighting_system}
\par The calculation of the ELPD requires that each stepwise model is estimated as many times as the number of time points ahead at which the predictive log-density must be calculated. This, in turn, increases the overall computational load and can be avoided using a faster, yet reliable, approximation of the ELPD, which is the \textit{Watanabe–Akaike Information Criterion}. With the WAIC the estimated effective number of parameters ($\hat{p}_{q}^{\text{waic}}$) is subtracted to the approximation of the Log-pointwise Predictive Density ($\widehat{lpd}_{q}$).
\begin{equation}
    \widehat{elpd}_{q}^{\text{waic}} = \widehat{lpd}_{q} - \hat{p}_{q}^{\text{waic}} 
    \label{eq:waic_formula}
\end{equation}
The $\widehat{lpd}_{q}$ is approximated by using draws from the posterior distribution computed with the whole event sequence and it represents an overestimate of the ELPD,
\begin{equation}
    \widehat{lpd}_{q}  = \sum_{i=L}^{M-A}{\log{\left(
    \frac{1}{B}\sum_{b=1}^{B}{p(e_{i+1},\ldots, e_{i+A}|E_{t_i},\bm{\beta}_{q}^{(b)})}   
    \right)}}
    \label{eq:lpd_approximation}
\end{equation}
where:
    \begin{itemize}
        \item $M$ is the number of observed relational events in the sequence;
        \item $L$ is a predefined minimum number of relational events from the ordered sequence that will be required to make predictions for future events in the WAIC;  
        \item $p(e_{i+1},\ldots, e_{i+A}|E_{t_i},\bm{\beta}_{q}^{(b)})$ is the posterior predictive density of the model $\mathcal{M}_q$ for A steps ahead predictions, that are $A$ future events occurring after the $i\text{-th}$ event and given the past history ($E_{t_i}$) of events until $t_i$, with $i=L,\ldots,(M-A)$; $\bm{\beta}_{q}^{(b)}$ is the $b\text{-th}$ vector of posterior draws for the $q\text{-th}$ stepwise model and it is generated from a multivariate normal distribution $\bm{\beta}_{q}\sim MVN(\hat{\bm{\beta}}_{q},\hat{\bm{\Sigma}}_{q})$ with parameters equal to the maximum likelihood estimates obtained by estimating the model over the whole sequence of events;
        \item $B$ is the number of posterior draws for each stepwise model.
    \end{itemize}
The effective number of parameters $\hat{p}_{q}^{\text{waic}}$ is calculated by using the same draws for the approximation of the $\widehat{lpd}_{q}$ and it consists in the sum of the sample variances of each prediction point,
\begin{equation}
    \hat{p}_{q}^{\text{waic}}  = \sum_{i=L}^{M-A}{\sum_{b=1}^{B}{\frac{1}{B-1}\left(\log{p(e_{i+1},\ldots, e_{i+A}|E_{t_i},\bm{\beta}_{q}^{(b)})}-\frac{1}{B}\sum_{b=1}^{B}{\log{p(e_{i+1},\ldots, e_{i+A}|E_{t_i},\bm{\beta}_{q}^{(b)})}}\right)^{2}}
    }
    \label{eq:effective_number_of_parameters_estimation}
\end{equation}
Hence, WAIC weights will be computed as
\begin{equation}
    w^{\text{WAIC}}_{q} = \frac{\exp{\left\lbrace\widehat{elpd}_{q}^{\text{waic}}\right\rbrace}}{\sum_{q=1}^{Q}{\exp{\left\lbrace\widehat{elpd}_{q}^{\text{waic}}\right\rbrace}}},\quad{}q=1,\ldots,Q
    \label{eq:waic_weights}
\end{equation}

\section{Case study: investigating the presence of memory decay in the sequence of demands sent among Indian socio-political actors}
\label{sec:case_study_investigating_the_presence_of_memory_in_the_demands_for_aid_between_indian_sociopolitical_actors}
We have now introduced our modeling approach, starting from a purely stepwise decay model to a continuous decay model based on model averaging of a set of stepwise models. In this section we illustrate the method by applying it to empirical data to provide insights about memory decay in contiuous time in a real-life social network. First, the empirical application and data are described. Next, the analyses are presented using different prespecified step-wise decay functions, followed by an application of the Bayesian model averaging estimated to obtain approximate smooth decay functions.

\subsection{Relational events between socio-political actors}
Data are retrieved from the ICEWS (Integrated Crisis Early Warning System) \cite{ICEWSrep}, which are available in the Harvard Dataverse repository. In ICEWS, relational events consist of interactions between socio-political actors which were extracted from news articles. Information as to the source actor, target actor and event type of the relational event are recorded along with geographical and temporal data that are available within the same article. Furthermore, event types are coded according to the CAMEO (Conflict and Mediation Event Observations) ontology.
In this analysis, we focus on the sequence of relational events within the country of India. Each event represented a request generated from an actor and targeted to another actor. Such requests range from humanitarian ones to military or economic ones and in this analysis this distinction isn't made. The specific relational event sequence used in our analysis is available upon request.

\par The event sequence includes M = 7567 dyadic events between June 2012 and April 2020 among the ten most active actor types: citizens, government, police, member of the Judiciary, India, Indian National Congress Party, Bharatiya Janata Party, ministry, education sector, and "other authorities". Since the time variable is recorded at a daily level, those events that occurred on the same day are considered evenly spaced throughout a day.

\par The network dynamics of interest are inertia, reciprocity, and transitivity closure which are included in the loglinear predictor of the dyadic event rates. Given a generic stepwise model with $K$ steps, the log-rate at any time $t\in[t_1,t_{M}]$ and for any request $e'$ in the risk set $\mathcal{R}$ (with $|\mathcal{R}|= N(N-1) = 90 $, i.e. all the possible dyads are included in the risk set) is:
\begin{equation}
    \begin{split}
        \log{\lambda(s_{e'},r_{e'},E_{t},\bm{\beta})} = \beta_0+ \sum_{k=1}^{K}{\beta_{\text{inertia}_k}\text{inertia}_k(s_{e'},r_{e'},t)}+
        \sum_{k=1}^{K}{\beta_{\text{reciprocity}_k}\text{reciprocity}_k(s_{e'},r_{e'},t)}+\\
        +\sum_{k=1}^{K}{\beta_{\text{transitivity closure}_k}\text{transitivity closure}_k(s_{e'},r_{e'},t)}
    \end{split}
    \label{eq:linear_predictor_real_data_application}
\end{equation}
where $\beta_{0}$ represents the logarithm of the baseline rate of requests and the remaining effects describe the estimated stepwise trends for the three network dynamics. Inertia quantifies the persistence of the sender in targeting its requests to the same receiver, for instance because the receiver is an actor with some socio-political relevance like a legal figure or authority. Reciprocity describes the level of reciprocation of the sender towards the receiver based on the past volume of interactions that the receiver addressed to the sender. Transitivity closure quantifies the level of information mediation by means of the volume of triads that can be potentially closed by the occurrence of event $e'$.

\subsection{Predefined stepwise decay models}
\begin{figure}[htp]
    \centering
    \includegraphics[scale=0.45]{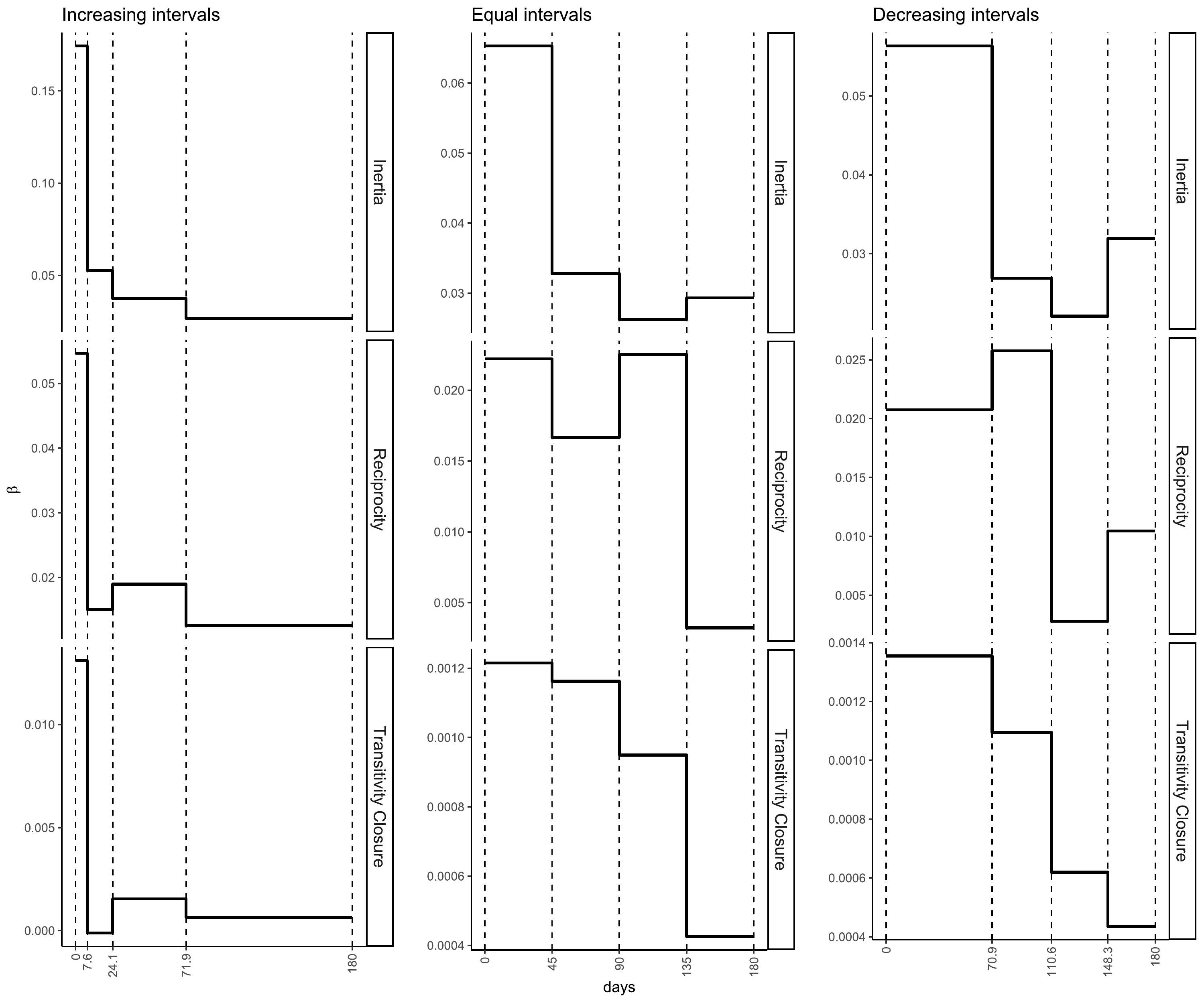}
  \caption{MLE estimates (by column) of three stepwise models (with $K = 4$), that are randomly chosen from the bag of the estimated models and each following one of the three interval types (increasing, equal and decreasing). The bold black line represents the stepwise function for each endogenous effect in the model and the vertical dashed lines indicate the time bounds characterizing the intervals. Therefore, each model follows different time widths.}
    \label{fig:predefined_stepwise_decay_models_example}
\end{figure}
\par As the maximum time ($\gamma_{K}$) that past events may affect current relational events we consider 180 days (roughly half an year). Furthermore three different predefined stepwise memory decay functions are considered by dividing the past in $K=4$ intervals with either increasing widths, equal widths, or decreasing widths, as described in Section \ref{subsec:defining a bag of stepwise relational event models}. Figure \ref{fig:predefined_stepwise_decay_models_example} shows the estimated stepwise decay functions for inertia, reciprocity, and transitivity given the three different interval configurations.

As expected all three models result in different estimated (discretized) shapes of memory decay. For instance, by comparing the estimates of the Transitivity Closure, we see that decreasing intervals and increasing intervals produces two contrasting decays where, not only both decays follow different shapes, but also the magnitude of the effect lies on different levels, lower for the decreasing intervals. One more comparison over the effect magnitude can be made between equal and decreasing intervals where trends evolve around the same magnitude, whereas for increasing intervals the maximum for the magnitude results to be more than twofold (Inertia and Reciprocity) if not tenfold (Transitivity Closure) the equal and the decreasing type. This discrepancy is not only given by the type of the intervals but is also generated by the time widths that characterize the stepwise models.

In sum, stepwise models having predefined interval configurations provide us with a very rough idea how fast memory decays in a given relational event network. On the other hand, predefined step-wise memory decay models do not provide insights about the shape of memory decay over the transpired time, or, for example, whether an approximated exponential decay is more likely than a approximated smooth one-step decrease. To learn this from an observed relational event network, we need the proposed weighting system for a bag of step-wise models together with a Bayesian model averaging approach.

\subsection{Approximately smooth memory decay models}
\label{sec:results}
\begin{figure}[htp]
    \centering
    \includegraphics[scale=0.5]{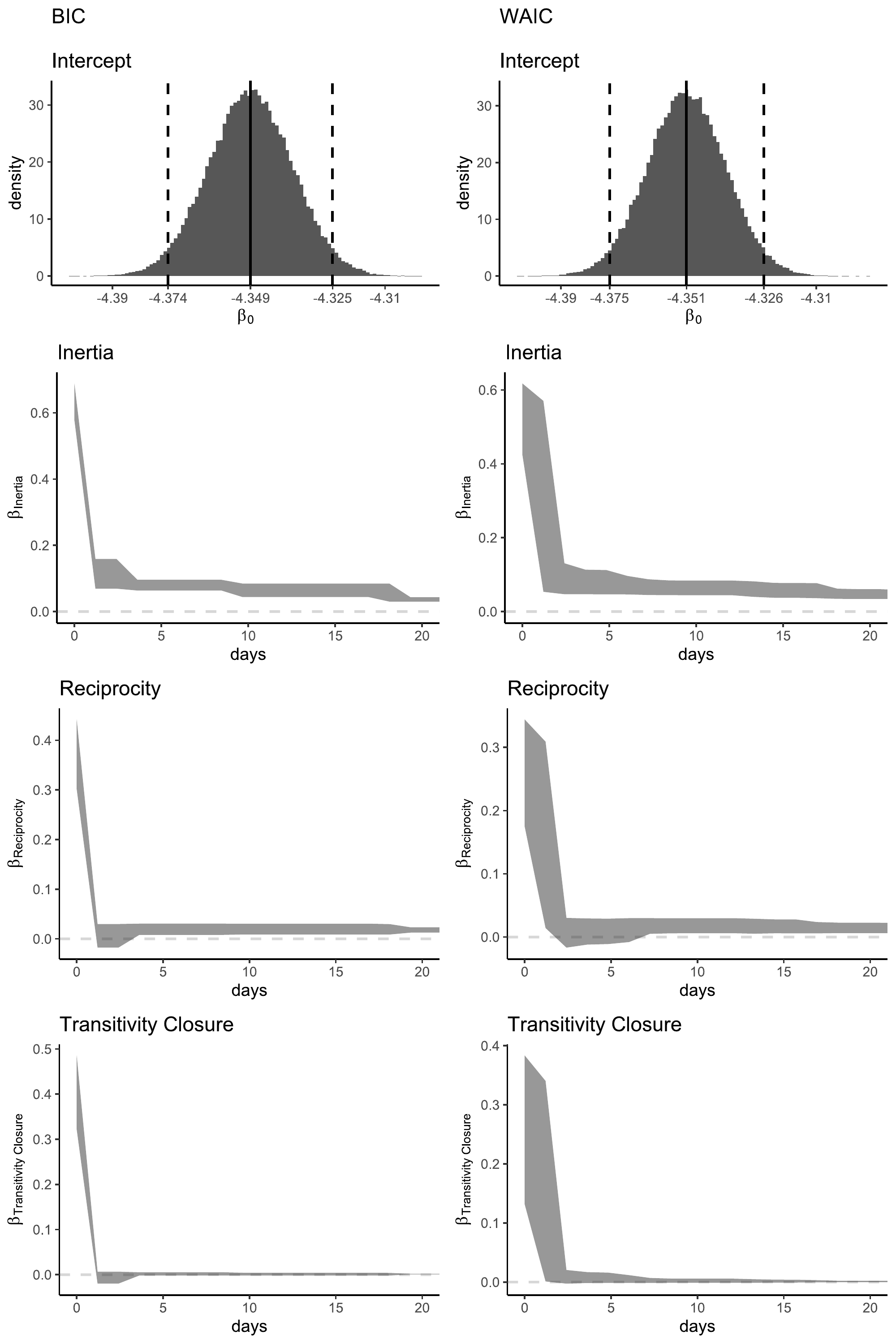}
  \caption{Posterior estimates resulting from the BMA with BIC (left) and WAIC (right) weights: posterior distribution for the intercept ($\beta_{0}$), posterior trends for inertia, reciprocity and transitivity closure. The gray area (dashed lines for the intercept) is generated by the posterior highest density intervals calculated until 20 days.}
    \label{fig:results_BMA_WAIC}
\end{figure}
\par For our bag of stepwise models (Section \ref{sec:A semi-parametric approach to estimate memory evolution}), three sets of 501 intervals were generated for $K=\left\lbrace3,4,5\right\rbrace$ steps (250 intervals with increasing sizes, 250 intervals with decreasing sizes, 1 with equal sizes). Thus in total, 1503 step-wise models were considered which will be used for the BMA stage where the posterior trend was estimated using the weighting systems presented in Section \ref{subsec:BIC_weighting_system} and in Section \ref{subsec:WAIC_weighting_system}. Figure \ref{fig:results_BMA_WAIC} shows the posterior trends resulting from two Bayesian Model Averaging approaches: one with BIC weights (left panels) and the other one with WAIC weights (right panels). Because most of the decay occurs in the first 20 days, only this period is plotted in the figure.
The intercept $\beta_{0}$ is the only parameter without a decay by definition and the posterior point estimate of the baseline event rate is $\exp{\left\lbrace\hat{\beta}_{0}\right\rbrace} \approx 0.0129$ (similar for both BIC and WAIC weights; upper panels).

Focusing on the results using the WAIC weights in Figure \ref{fig:results_BMA_WAIC}, all the three trends show a clear approximately exponential memory decay (right panels). This drastic decrease near zero suggests that the most recent requests have a much higher impact on the event rate than the less recent ones. Therefore, the trend observed for inertia suggests the presence of a persistence in actors that tend to keep sending requests to the same recipient of their more recent requests, showing inertia based on the requests that happened in a fairly recent past, rather than inertia based on requests over a longer period of time.

\par For reciprocity, we see that the effect drops a bit faster than inertia and stabilizes around a low value that decreases further, indicating that actors reciprocate on requests received in the very recent past, but requests that were not responded to quickly are soon forgotten and are unlikely to be followed up. Norms of reciprocity are clearly not strong and non-reciprocated requests disappear from social memory very quickly.
Finally, transitivity effect vanishes similarly fast. This can be explained as follows. Considering that dyadic requests only briefly trigger the tendency to respond, it makes sense that common communication partners also have only very short-lived influence on future interaction between two parties. 

\par The resulting trends obtained by using the BIC weights approximately follow the same decays as found with WAIC. However, we see an approximate stepwise trend using BIC weights. This can be explained by the fact that the BIC becomes increasingly large for the stepwise model that is closest to the true (smooth) model (in terms of Kullback-Leibler distance \cite{grunwald2017}). Thus the weight of the stepwise model that is closest to the true smooth model dominates over the weights of all other stepwise models. This illustrates that the BIC is useful for finding the best fitting stepwise model, which, in this case, has increasing interval widths over the transpired time having a roughly exponential decay. On the other hand, the BIC is not useful for finding an approximate smooth decay trend. For this purpose the WAIC is recommended. 


\section{Discussion}
\label{sec:discussion}
\par In this paper we presented different methods for learning how past interactions between social actors affect future interactions in the network. First a $K$-stepwise model was considered where memory decay about past interactions was approximated using a discrete stepwise trend. This model can be estimated using existing software functions for relational event analysis. The proposed Bayesian model averaged memory decay estimator will be made available in a new R package in the coming month.

As memory naturally decays in a smooth continuous trend, the next key contribution was a novel Bayesian model averaging approach to estimate the memory decay in a relational modeling framework where events are assumed to continuously change their influence on the network dynamics (and, in turn, on the event rate) according to their recency. 
The promising aspect of this semi-parametric approach lies on its ability to learn the shape of the memory decay without making any parametric assumption about it. Furthermore by building on the stepwise model, the proposed method is computationally feasible. Two different weighting systems were considered for Bayesian model averaging of a bag of stepwise models: The BIC and the WAIC. As was illustrated the BIC is useful for finding \textit{the} best fitting stepwise model for a given empirical relational event history. The BIC however is not suitable for finding an approximate smooth trend of memory decay, as all the weights are placed on the single stepwise model that is closest to the true smooth decay model. This issue is not present when using the WAIC as the Bayesian model average of many stepwise models results in a smooth trend of the decay of the actors' memory about past events in social networks.

Two advancements of the method will consist in extending it to different event types (sentiments) and to other endogenous statistics that measure specific network dynamics. The former will be helpful for the understanding of the differences in memory decays given different sentiments characterizing the interactions. For instance, one expects negative events (e.g., a country threatening another country, a pupil insulting a peer, a teacher rebuking a student) to have a memory decay that is slower and more persistent than the one observed for positive events (e.g., a teacher praising a student, a country cooperating with another country). This difference may apply as well to other event type settings where possible different memory shapes can emerge. The other advancement will consist in first formulating new endogenous statistics that measure specific behavioral patterns in the network, then in estimating and interpreting the decay of their effect.

Finally, future investigations could also focus on formulating and estimating key characteristics that describe the memory decay as well as on testing differences of such features across either sentiments of events or groups of actors. We expect that the acquired ability of both estimating and testing on parameters that describe this memory process is a crucial step towards a more accurate understanding of network dynamics developing at a local as well as at a global level.


\section*{Author biographies}
\par \textbf{Giuseppe Arena} graduated from the University of Padova with a Master degree in Statistics. He is now pursuing a PhD at the Department of Methodology and Statistics at Tilburg University. His main research interests are social network analysis, Bayesian inference and analysis of memory retention process in relational event data.
\par \textbf{Joris Mulder} is an associate professor in the Department of Methodology and Statistics at Tilburg University. He holds a PhD in applied Bayesian statistics from Utrecht University. His research focuses on Bayesian model selection and social network
modeling.
\par \textbf{Roger Th. A. J. Leenders} is a professor at the Jheronimus Academy of Data Science and in the Department of Organization Studies at Tilburg University. He holds a PhD in sociology from the University of Groningen. He has published broadly on social network analysis, teams, innovation, and organization behavior in leading journals such as Organization Science, the Journal of Applied Psychology, the Journal of Product Innovation Management, Social Networks, and the Academy of Management Journal.
\newpage
\appendix
\section{Appendix}
\subsection{Endogenous statistics}
\label{appendix:table_with_endogenous_statistics}
\par In Table \ref{table:first_and_second_order_endogenous_statistics} the indicator variable for any event $e$ where $\left(s_e=i,r_e=j\right)$ follows the short notation $\mathbb{I}_e(i,j)$ and the same applies to any other dyad. Given each statistic, the formula in the first row shows the interval definition of the statistic as regards dyad $(i,j)$ in the $k\text{-th}$ interval; whereas, the formula in the second row shows the continuous definition where $\beta(\gamma,\bm{\theta})$ is the trend function that follows one of the decays discussed in Section \ref{sec:the_continuous_nature_of_memory_decay} or another more complex evolution. Note how in the continuous formulas the event history at $t_m$, that is $E-{t_{m-1}}$, doesn't depend on any interval.
\begin{table}
\begin{center}
    \scalebox{0.83}{
    \begin{tabular}{|c|c|l|}
    \hline
    &\textbf{Endogenous Statistic}&\multicolumn{1}{c|}{\textbf{Formula}} \\ \cline{2-3}
          \parbox[t]{2mm}{\multirow{6}{*}[-17em]{\rotatebox[origin=c]{90}{\textbf{First Order}}}} && \\
&    \begin{tabular}{c}
    Inertia \\
    \begin{tikzpicture}[scale=0.50]
    \node[state,scale=0.50] (x) at (0,0) {$i$};
    \node[state,scale=0.50] (y) [right =of x] {$j$};
    \path[scale=0.50] (x) edge (y);
    \path[gray,dashed,scale=0.50] (x) edge[bend right,scale=0.50] (y);
    \addvmargin{2mm}
    \end{tikzpicture}
    \end{tabular}
    &
    \begin{tabular}{l}
        $\text{inertia}_{k}(i,j,t_{m}) = \sum_{e\in E_{t_{m-1},k}}^{}{\mathbb{I}_e(i,j)}$  \\[0.5em]
        $\text{inertia}(i,j,t_{m},\beta(\gamma,\bm{\theta})) = \sum_{e\in E_{t_{m-1}}}^{}{\mathbb{I}_e(i,j)\beta(\gamma_{e}(t_m),\bm{\theta})}$
    \end{tabular} \\ \cline{2-3}
&    \begin{tabular}{c}
    Reciprocity \\
    \begin{tikzpicture}[scale=0.50]
        \node[state,scale=0.50] (x) at (0,0) {$i$};
        \node[state,scale=0.50] (y) [right =of x] {$j$};
        \path[scale=0.50] (x) edge (y);
        \path[gray,dashed,scale=0.50] (y) edge[bend left,scale=0.50] (x);
        \addvmargin{2mm}
    \end{tikzpicture}
    \end{tabular} 
    &
    \begin{tabular}{l}
        $\text{reciprocity}_{k}(i,j,t_{m}) = \sum_{e\in E_{t_{m-1},k}}^{}{\mathbb{I}_e(j,i)}$ \\[0.5em] 
        $\text{reciprocity}(i,j,t_{m},\beta(\gamma,\bm{\theta})) = \sum_{e\in E_{t_{m-1}}}^{}{\mathbb{I}_e(j,i)\beta(\gamma_{e}(t_m),\bm{\theta})}$
    \end{tabular} \\ \cline{2-3} 
&    \begin{tabular}{c}
    Sender  in-degree \\
    \begin{tikzpicture}[scale=0.50]
        \node[state,scale=0.50] (x) at (1,-1) {$i$};
        \node[state,scale=0.50] (y) [right =of x] {$j$};
        \node[state,scale=0.50] (a) at (-1.5,1) {$a$};
        \node[state,scale=0.50] (b) at (-1.5,0) {$b$};
        \node[scale=0.50] (s) at (-1.5,-1) {$\vdots$};
        \node[state,scale=0.50] (z) at (-1.5,-2) {$z$};
        \path[scale=0.50] (x) edge (y);
        \path[gray,dashed,scale=0.50] (y) edge[bend left,scale=0.50] (x);
        \path[gray,dashed,scale=0.50] (a) edge[bend left,scale=0.50] (x);
        \path[gray,dashed,scale=0.50] (b) edge[bend left,scale=0.50] (x);
        \path[gray,dashed,scale=0.50] (z) edge[bend right,scale=0.50] (x);
        \addvmargin{2mm}
    \end{tikzpicture}
    \end{tabular}
    &
    \begin{tabular}{l}
        $\text{indegree}^{\text{snd}}_{k}(i,j,t_{m}) = \sum_{l\in\mathcal{S}\setminus\left\lbrace i\right\rbrace}^{}{\sum_{e\in E_{t_{m-1},k}}^{}{\mathbb{I}_e(l,i)}}$ \\[1.5em]
        $\text{indegree}^{\text{snd}}(i,j,t_{m},\beta(\gamma,\bm{\theta})) =\sum_{l\in\mathcal{S}\setminus\left\lbrace i\right\rbrace}^{}{\sum_{e\in E_{t_{m-1}}}^{}{\mathbb{I}_e(l,i)\beta(\gamma_{e}(t_m),\bm{\theta})}}$
    \end{tabular} \\ \cline{2-3}
&    \begin{tabular}{c}
    Sender   out-degree \\
    \begin{tikzpicture}[scale=0.50]
            \node[state,scale=0.50] (x) at (1,-1) {$i$};
            \node[state,scale=0.50] (y) [right =of x] {$j$};
            \node[state,scale=0.50] (a) at (-1.5,1) {$a$};
            \node[state,scale=0.50] (b) at (-1.5,0) {$b$};
            \node[scale=0.50] (s) at (-1.5,-1) {$\vdots$};
            \node[state,scale=0.50] (z) at (-1.5,-2) {$z$};
            \path[scale=0.50] (x) edge (y);
            \path[gray,dashed,scale=0.50] (x) edge[bend right,scale=0.50] (y);
            \path[gray,dashed,scale=0.50] (x) edge[bend right,scale=0.50] (a);
            \path[gray,dashed,scale=0.50] (x) edge[bend right,scale=0.50] (b);
            \path[gray,dashed,scale=0.50] (x) edge[bend left,scale=0.50] (z);
            \addvmargin{2mm}
        \end{tikzpicture}
    \end{tabular}
    &
    \begin{tabular}{l}
        $\text{outdegree}^{\text{snd}}_{k}(i,j,t_{m}) = \sum_{l\in\mathcal{S}\setminus\left\lbrace i\right\rbrace}^{}{\sum_{e\in E_{t_{m-1},k}}^{}{\mathbb{I}_e(i,l)}}$ \\[1.5em]
         $\text{outdegree}^{\text{snd}}(i,j,t_{m},\beta(\gamma,\bm{\theta})) = \sum_{l\in\mathcal{S}\setminus\left\lbrace i\right\rbrace}^{}{\sum_{e\in E_{t_{m-1}}}^{}{\mathbb{I}_e(i,l)\beta(\gamma_{e}(t_m),\bm{\theta})}}$
    \end{tabular}    
      \\ \cline{2-3}
&        \begin{tabular}{c}
        Receiver   in-degree \\
         \begin{tikzpicture}[scale=0.50]
            \node[state,scale=0.50] (x) at (-2.5,-1) {$i$};
            \node[state,scale=0.50] (y) [right =of x] {$j$};
            \node[state,scale=0.50] (a) at (2.5,1) {$a$};
            \node[state,scale=0.50] (b) at (2.5,0) {$b$};
            \node[scale=0.50] (s) at (2.5,-1) {$\vdots$};
            \node[state,scale=0.50] (z) at (2.5,-2) {$z$};
            \path[scale=0.50] (x) edge (y);
            \path[gray,dashed,scale=0.50] (x) edge[bend right,scale=0.50] (y);
            \path[gray,dashed,scale=0.50] (a) edge[bend right,scale=0.50] (y);
            \path[gray,dashed,scale=0.50] (b) edge[bend right,scale=0.50] (y);
            \path[gray,dashed,scale=0.50] (z) edge[bend left,scale=0.50] (y);
            \addvmargin{2mm}
        \end{tikzpicture}
        \end{tabular}
        &
        \begin{tabular}{l}
        $\text{indegree}^{\text{rec}}_{k}(i,j,t_{m}) = \sum_{l\in\mathcal{S}\setminus\left\lbrace j\right\rbrace}^{}{\sum_{e\in E_{t_{m-1},k}}^{}{\mathbb{I}_e(l,j)}}$ \\[1.5em]
        $\text{indegree}^{\text{rec}}(i,j,t_{m},\beta(\gamma,\bm{\theta})) = \sum_{l\in\mathcal{S}\setminus\left\lbrace j\right\rbrace}^{}{\sum_{e\in E_{t_{m-1}}}^{}{\mathbb{I}_e(l,j)\beta(\gamma_{e}(t_m),\bm{\theta})}}$
        \end{tabular}
        \\ \cline{2-3}   
&        \begin{tabular}{c}
        Receiver  out-degree \\
        \begin{tikzpicture}[scale=0.50]
            \node[state,scale=0.50] (x) at (-2.5,-1) {$i$};
            \node[state,scale=0.50] (y) [right =of x] {$j$};
            \node[state,scale=0.50] (a) at (2.5,1) {$a$};
            \node[state,scale=0.50] (b) at (2.5,0) {$b$};
            \node[scale=0.50] (s) at (2.5,-1) {$\vdots$};
            \node[circle,state,scale=0.50] (z) at (2.5,-2) {$z$};
            \path[scale=0.50] (x) edge (y);
            \path[gray,dashed,scale=0.50] (y) edge[bend left,scale=0.50] (x);
            \path[gray,dashed,scale=0.50] (y) edge[bend left,scale=0.50] (a);
            \path[gray,dashed,scale=0.50] (y) edge[bend left,scale=0.50] (b);
            \path[gray,dashed,scale=0.50] (y) edge[bend right,scale=0.50] (z);
            \addvmargin{2mm}
        \end{tikzpicture}
        \end{tabular}
        &
        \begin{tabular}{l}
            $\text{outdegree}^{\text{rec}}_{k}(i,j,t_{m}) = \sum_{l\in\mathcal{S}\setminus\left\lbrace j\right\rbrace}^{}{\sum_{e\in E_{t_{m-1},k}}^{}{\mathbb{I}_e(j,l)}}$ \\[1.5em] $\text{outdegree}^{\text{rec}}(i,j,t_{m},\beta(\gamma,\bm{\theta})) = \sum_{l\in\mathcal{S}\setminus\left\lbrace j\right\rbrace}^{}{\sum_{e\in E_{t_{m-1}}}^{}{\mathbb{I}_e(j,l)\beta(\gamma_{e}(t_m),\bm{\theta})}}$ 
        \end{tabular}
        \\ \hline
\parbox[t]{2mm}{\multirow{2}{*}[-6em]{\rotatebox[origin=b]{90}{\textbf{Second Order}}}} && \\
&    \begin{tabular}{c}
    Transitivity closure \\
    \begin{tikzpicture}[scale=0.50]
            \node[state,scale=0.50] (x) at (0,2) {$i$};
            \node[state,scale=0.50] (y) at (-1.5,1) {$l$};
            \node[state,scale=0.50] (z) at (0,0) {$j$};
            \path[gray,dashed,scale=0.50] (x) edge (y);
            \path[scale=0.50] (x) edge[bend left,scale=0.50] (z);
            \path[gray,dashed,scale=0.50] (y) edge (z);
            \addvmargin{2mm}
    \end{tikzpicture} 
    \end{tabular}
    &
    \begin{tabular}{l}
        $\displaystyle \text{transitivity closure}_{k}(i,j,t_{m}) =$ \\ $
        \sum_{l\in\mathcal{S}\setminus\left\lbrace i,j\right\rbrace}{
        \sum_{\substack{e\in E_{t_{m-1},k}}}{\sum_{\substack{e^{*}\in E_{t_{m-1}}:\\ t_{e^{*}}\in[t_{e}-\gamma_{e}(t_m),t_{e})}}{\mathbb{I}_e(l,j)}\mathbb{I}_{e^{*}}(i,l)}}$ \\[1.5em]
       $\text{transitivity closure}(i,j,t_{m},\beta(\gamma,\bm{\theta})) = $ \\ $
        \sum_{l\in\mathcal{S}\setminus\left\lbrace i,j\right\rbrace}{
       \sum_{\substack{e\in E_{t_{m-1}}}}{\sum_{\substack{e^{*}\in E_{t_{m-1}}:\\ t_{e^{*}}\in[t_{e}-\gamma_{e}(t_m),t_{e})}}{\mathbb{I}_e(l,j)}\mathbb{I}_{e^{*}}(i,l)\beta(\gamma_{e}(t_m),\bm{\theta})}}$ 
       \\[2em]
     \end{tabular}
     \\ \cline{2-3}
&    \begin{tabular}{c}
    Cyclic closure \\
    \begin{tikzpicture}[scale=0.50]
            \node[state,scale=0.50] (x) at (0,2) {$i$};
            \node[state,scale=0.50] (y) at (-1.5,1) {$l$};
            \node[state,scale=0.50] (z) at (0,0) {$j$};
            \path[gray,dashed,scale=0.50] (y) edge (x);
            \path[scale=0.50] (x) edge[bend left] (z);
            \path[gray,dashed,scale=0.50] (z) edge (y);
            \addvmargin{2mm}
    \end{tikzpicture}
    \end{tabular}
    &
    \begin{tabular}{l}
        \\[0.5em]
        $\displaystyle \text{cyclic closure}_{k}(i,j,t_{m}) = $ \\ $
        \sum_{l\in\mathcal{S}\setminus\left\lbrace i,j\right\rbrace}{
        \sum_{\substack{e\in E_{t_{m-1},k}}}{\sum_{\substack{e^{*}\in E_{t_{m-1}}:\\ t_{e^{*}}\in[t_{e}-\gamma_{e}(t_m),t_{e})}}{\mathbb{I}_e(l,i)}\mathbb{I}_{e^{*}}(j,l)}}$ \\[1.5em]
       $\text{cyclic closure}(i,j,t_{m},\beta(\gamma,\bm{\theta})) = $ \\ $
        \sum_{l\in\mathcal{S}\setminus\left\lbrace i,j\right\rbrace}{
       \sum_{\substack{e\in E_{t_{m-1}}}}{\sum_{\substack{e^{*}\in E_{t_{m-1}}:\\ t_{e^{*}}\in[t_{e}-\gamma_{e}(t_m),t_{e})}}{\mathbb{I}_e(l,i)}\mathbb{I}_{e^{*}}(j,l)\beta(\gamma_{e}(t_m),\bm{\theta})}}$
       \\[2em]
     \end{tabular}
       \\ \hline
    \end{tabular}}
    \end{center}
    \caption{First and second order endogenous statistics: the formula in the first row is the interval definition of the statistic, whereas the formula in the second row represents the continuous definition of the statistic where the function $\beta(\gamma,\bm{\theta})$ describes the decay of the effect.}
    \label{table:first_and_second_order_endogenous_statistics}
\end{table}
\subsection{From stepwise to continuous effects}
\label{appendix:from_stepwise_to_continuous_effects}
\par
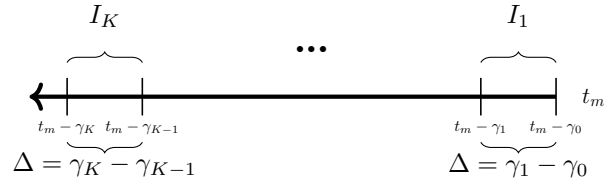
\begin{figure}[!h]
    \begin{center}
        \begin{tikzpicture}
        \foreach \x/\y in {0.5/, 1.5/, 6/, 7/ }
         \draw[-,semithick] (\x,0.25) -- (\x,-0.25) node[below]{$\y$};
        \node[align=right, above, scale=0.6] at (-0.3,-0.2) 
        {};
        \node[align=right, above, scale=0.6] at (7.5,-0.2) 
        {\Large$t_{m}$};
        \node[align=right, below, scale=0.6] at (7,-.25) 
        {$t_{m} - \gamma_0$};
        \node[align=right, below, scale=0.6] at (6,-.25) 
        {$t_{m} - \gamma_1$};
        \node[align=right, below, scale=0.6] at (0.5,-.25) 
        {$t_{m} - \gamma_K$};
        \node[align=right, below, scale=0.6] at (1.5,-.25) 
        {$t_{m} -\gamma_{K-1}$};
        \node[align=right, below, scale=0.6] at (3.75,0.75)
        {$\bullet\bullet\bullet$};
        \draw [-,decorate,decoration={brace,amplitude=4pt,raise=3ex}]
        (6,0) -- (7,0) node[midway,yshift=2.3em]{$I_1$};
        \draw[-,decorate,decoration={brace,amplitude=4pt,raise=4ex,mirror}]
        (6,0) -- (7,0) node[midway,yshift=-3.4em]{$\Delta = \gamma_{1}-\gamma_{0}$};
        \draw[-,decorate,decoration={brace,amplitude=4pt,raise=3ex}]
        (0.5,0) -- (1.5,0) node[midway,yshift=2.3em]{$I_K$};
        \draw[-,decorate,decoration={brace,amplitude=4pt,raise=4ex,mirror}]
        (0.5,0) -- (1.5,0) node[midway,yshift=-3.4em]{$\Delta = \gamma_{K}-\gamma_{K-1}$}; 
        \draw[<-, ultra
        thick] (0,0)--(7,0);
        \end{tikzpicture}
        \end{center}
        \caption{$K$ evenly spaced intervals, with time widths $\bm{\gamma} = (\gamma_0, \gamma_1, \ldots, \gamma_K)$ such that $\gamma_k-\gamma_{k-1} = \Delta$ for $k=1,\ldots,K$.}
        \label{fig:from_stepwise_to_continuous_effects}
\end{figure}

 Consider an increasing sequence of $K+1$ time widths $\bm{\gamma} = (\gamma_0, \gamma_1, \ldots, \gamma_K)$, such that $\gamma_k-\gamma_{k-1} = \Delta$ for $k=1,\ldots,K$ (i.e. evenly spaced intervals). A graphical representation of intervals at $t_{m}$ is presented in Figure \ref{fig:from_stepwise_to_continuous_effects}.

In the context of endogenous statistics that are defined on intervals (see Section \ref{sec:A stepwise memory decay model:first order endogenous statistics}), one could already apply the formulas in Appendix \ref{appendix:table_with_endogenous_statistics} and then estimate the stepwise trend for each network statistic of interest. In general, when intervals are evenly spaced, we could write $\gamma_k =
k\cdot\frac{\gamma_{K}}{K}$ for $k=0,\ldots,K$, where $\gamma_{K}$ is the
largest observable width (it can be the length of the study itself). If the number of intervals ($K$) increases, their size ($\Delta$), in turn, shrinks. Indeed, considering the size of an interval that is calculated as the difference between two adjacent widths, $\Delta = \left(\gamma_k-\gamma_{k-1}\right)$.
$$
\lim_{K\rightarrow\infty}{\left(\gamma_k-\gamma_{k-1}\right)}=
\lim_{K\rightarrow\infty}{\left[ k\cdot\frac{\gamma_{max}}{K}-(k-1)\cdot\frac{\gamma_{max}}{K}\right]}=
\lim_{K\rightarrow\infty}{\frac{\gamma_{max}}{K}}=0
\label{from_few_intervals_to_many_intervals_limit}
$$
This result holds for $k=1,\ldots,K$. Therefore, an extreme scenario consists in a large number of intervals whose sizes are so small that at $t_m$ each of them contains only one or no relational event. As a consequence of this, one would estimate a stepwise trend where each step is defined approximately on a value of the transpired time and it represents the relative effect based on those events that assumed that specific value throughout the event histories ($E_{t}$, with $t=t_{1},\ldots,t_{M}$). Indeed, any event since its occurrence assumes a value reflecting its recency which is updated at every time point onward and, thus, it increases over time (from $t_{1}$ to $t_{M}$ if considering the time points where events were observed). Every value of transpired time calculated at each time point can be observed at least once in the network and when it is observed multiple times this happens at different time points. For instance, two different events could both occur 33 minutes earlier than the present time point but with the condition that the present time point they refer to is different for both of them (because events are assumed not to occur at the same time point). Finally, the estimation of the effects over such a large number of intervals is impractical and it serves only to convey insights about the possibility of continuously changing effects in contrast to stepwise decays. Continuous effects imply a more realistic view of the effect dynamics in which events do not assume a constant effect but this changes with their recency, whereas stepwise effects suppose the less realistic scenario where effects are assumed to be constant within intervals and this also entails that estimates significantly depend both on the number and on the size of the intervals.
\newpage
\subsection{Interval generator (the algorithm)}
\label{appendix:intervals_generator}
\begin{algorithm}[!h]\small
    \caption{Generating $S$ intervals with $K$ steps (having either increasing or decreasing sizes).}
    \Set{(inputs and memory allocation for the output)}:\ 
    \begin{itemize}
    	\item[] $K\leftarrow$ number of steps for each sequence
    	\item[] $S\leftarrow$ number of sequences to generate
    	\item[] $s=1 \leftarrow$ starting with generating the first sequence
    	\item[] $\text{min\_size}\leftarrow$ minimum size intervals
    	\item[] $\gamma_K \leftarrow$ maximum time width that each sequence can reach
    	\item[] $decreasing \leftarrow$ logical TRUE/FALSE whether to generate either decreasing sizes or increasing sizes intervals
    	\item[] $\mathcal{W}\leftarrow$ empty matrix of dimensions $[S\times K]$ where to store the generated sequences of widths $\bm{\gamma}=(\gamma_1,\ldots,\gamma_K)$ (excluding $\gamma_0$ which is equal to zero by default)
    \end{itemize}
	 \While{s $\le$ S}{
		\BlankLine
		\Generate{} $\bm{\xi}\sim Dir(K,\bm{\alpha})$ with $\bm{\alpha}=\bm{1}_{K}$\;
		\BlankLine
		\Sort{ascending} $\bm{\xi}$\;
		\BlankLine
		\myWhile{$\min{\left\lbrace\bm{\xi}\right\rbrace}<\text{min\_size}$}{
			\BlankLine
			\Generate{} $\bm{\xi}\sim Dir(K,\bm{\alpha})$ with $\bm{\alpha}=\bm{1}_{K}$\;
			\BlankLine
			\Sort{ascending} $\bm{\xi}$\;
		}
		\BlankLine
		\myif{decreasing = TRUE} \mythen{} \Sort{descending} $\bm{\xi}$\;
		\BlankLine
		$\bm{\gamma}\leftarrow$ \Cumsum{} of $\bm{\xi}$\; 
		\BlankLine
		\Update{} $\bm{\gamma} = \bm{\gamma} * \gamma_K$\;
		\BlankLine
		\Save{} $\bm{\gamma}$ in the $s\text{-th}$ row of $\mathcal{W}$\;
		\Update{} $s = s + 1$\;
	}
	\BlankLine
	\Return{}$\mathcal{W}$\;
\end{algorithm}
    



\end{document}